\documentclass[twocolumn]{emulateapj}
\pdfoutput=1 
\usepackage{graphicx}
\usepackage{natbib}
\usepackage{multirow}
\usepackage{amsmath}
\usepackage{amssymb}
\usepackage{rotating}

\newcommand{\hii}{HII }

\newcommand{\csiso}{$^{13}$CS}
\newcommand{\methcy}{CH$_3$CN}
\newcommand{\methacet}{CH$_3$CCH}
\newcommand{\cyano}{HC$_3$N}
\newcommand{\hcniso}{H$^{13}$CN}
\newcommand{\hoco}{HOCO$^+$}
\newcommand{\hco}{HCO$^+$}
\newcommand{\diaz}{N$_2$H$^+$}
\newcommand{\hcoiso}{H$^{13}$CO$^+$}
\newcommand{\hnciso}{HN$^{13}$C}
\newcommand{\cthree}{c-C$_3$H$_2$}
\newcommand{\ctwo}{C$_2$H}
\newcommand{\msun}{M$_{\odot}$}
\newcommand{\cm}{cm$^{-3}$}
\newcommand{\kms}{km s$^{-1}$}

\begin{document}
\title{Origins of Scatter in the Relationship Between HCN 1-0 and Dense Gas Mass in the Galactic Center}
\author{Elisabeth A.C. Mills}
\affil{San Jose State University, 1 Washington Square, San Jose, CA 95192, USA}
\email{elisabeth.mills@sjsu.edu}

\author{Cara Battersby}
\affil{Harvard-Smithsonian Center for Astrophysics, 60 Garden St., Cambridge, MA 02138, USA}

\begin{abstract}

We investigate the correlation of HCN 1-0 with gas mass in the central 300 pc of the Galaxy. We find that on the $\sim$ 10 pc size scale of individual cloud cores, HCN 1-0 is well correlated with dense gas mass when plotted as a log-log relationship. There is $\sim$0.75 dex of scatter in this relationship from clouds like Sgr B2, which has an integrated HCN 1-0 intensity of a cloud less than half its mass, and others that have HCN 1-0 enhanced by a factor of 2-3 relative to clouds of comparable mass. We identify the two primary sources of scatter to be self-absorption and variations in HCN abundance. We also find that the extended HCN 1-0 emission is more intense per unit mass than in individual cloud cores. In fact the majority (80\%) of HCN 1-0 emission comes from extended gas with column densities below $7 \times10^{22}$ cm$^{-2}$, accounting for 68\% of the total mass. We find variations in the brightness of HCN 1-0 would only yield a $\sim 10\%$ error in the dense gas mass inferred from this line in the Galactic center. However, the observed order of magnitude HCN abundance variations, and the systematic nature of these variations, warn of potential biases in the use of HCN as dense gas mass tracer in more extreme environments such as AGN and shock-dominated regions. We also investigate other 3 mm tracers, finding that HNCO is better correlated with mass than HCN, and might be a better tracer of cloud mass in this environment.

\end{abstract}

\section{Introduction}

While the history of the cosmic star formation rate is well known, peaking around a redshift of 1.9 \citep{Madau98, MD14}, the physics that drives the variation of the star formation rate with redshift, including its steep decline since a redshift of 2 \citep{Behroozi13,Moster13}, are still uncertain. In particular, there has been a lack of direct measurements of molecular gas mass and density over the same timescales to connect the star formation properties with the conditions in the gas reservoir. This lack of data is largely due to the difficulties of observing molecular gas, especially at large distances, which has previously limited observations to bright tracers of molecular gas (e.g., single lines of CO or HCN) that serve as proxies for (dense) gas mass. The bulk of these measurements have been limited to nearby systems \citep[z $<$0.1;][]{GS04a,GC08,Leroy09,Saint11,GB12}, though recently larger samples have begun to push out beyond a redshift of 1 \citep[e.g.,][]{Gao07,Tacc10,Tacc13,Saint13,CW13,Genzel15}.

In the local universe, the use of CO as a gas tracer reveals a fundamental relationship between star formation (as traced by far-infrared luminosity) and gas surface density \citep{Kennicutt98}, although variations in the abundance and excitation of CO can change the CO to gas mass conversion factor \citep{Sandstrom13,Mashian15}. However, in massive, infrared-luminous galaxies, CO 1-0 is systematically weak compared to the far-infrared luminosity, and the Schmidt-Kennicutt relationship is not linear \citep{GS04a}. In contrast, HCN 1-0 retains a linear relationship with far-infrared luminosity over 7-8 orders of magnitude in infrared luminosity \citep[The Gao-Solomon relation][]{GS04b,Wu05}. The difference in behavior between HCN 1-0 and CO 1-0 in these systems can be explained if the HCN:CO ratio represents the fraction of dense (n $\gtrsim\ 10^4$ \cm) gas, which is suggested to be more fundamentally relevant for star formation \citep{Lada12}. 

In recent years the universal applicability of the Gao-Solomon relation to all extragalactic environments has been called into question. HCN 1-0 is observed to be anomalously bright in active galactic nuclei (AGN), compared to CO and \hco\, \citep{Davies,Imanishi09,Kohno03}. \cite{GC06} also find that HCN 1-0 may be overluminous compared to \hco\, in ultraluminous infrared galaxies (ULIRGs), and \cite{Privon15} also find that overluminous HCN 1-0 is not limited to galaxies containing AGN. In addition to these indications that HCN 1-0 is overluminous in some environments, there are observations that a single power-law spectrum is insufficient to describe both normal galaxies and ULIRGs. Other studies have found that the Gao-Solomon relation becomes superlinear, with ULIRGs having a systematically higher (by a factor of 2-3) ratio of far-infrared luminosity compared to HCN 1-0, which is interpreted as increased star formation efficiency in these systems \citep{GC08,GB12}. \cite{Gao07} similarly found that their sample of high-redshift galaxies had systematically brighter L$_{FIR}$ compared to HCN 1-0. It is standard to explain the variations in the HCN:IR ratio as differences in the star formation efficiency (or gas depletion time) as a function of environment, rather than a failure of HCN 1-0 to trace the total dense gas mass \citep[e.g.,][]{Juneau09,Leroy15,Usero15}.

However, there are several reasons that HCN 1-0 could fail to be a reliable tracer of the total dense gas mass in all environments. Although the modeling of \cite{GS04b} indicate that HCN 1-0 intensity should be proportional to dense gas mass across a wide range of physical conditions, this assumes both a constant abundance and purely collisional excitation.  X-ray and photon-dominated regions are suggested to change the gas chemistry, increasing the abundance of HCN relative to other species \citep{LD96,Meijerink07,Harada13}, while a strong 14 $\mu$m radiation field can induce radiative pumping in HCN, altering its level populations \citep{ZT86,Aalt07a,Sakamoto10,Mills13}. Infrared pumping has also been suggested to induce weak masing in the 1-0 line \citep{Matsushita15}, analogous to stronger masers observed in the envelopes of asymptotic giant branch stars \citep{Olof93,Izumiura95}. Although the optical thickness of HCN 1-0 (or CO 1-0) is suggested not to matter significantly, as extragalactic scaling relationships are often explained as counting optically thick (or thin) clouds, which as an ensemble are then proportional to the total mass \citep[e.g.][]{Wu05}, this assumption can break down: for example, when HCN is subject to self-absorption in very high column density environments like those in the nuclei of ULIRGs \citep{Aalto15b}. 

To understand the relationship between the molecular gas supply and the peak and decline of the cosmic star formation rate, we need to be able to make accurate measurements of the gas mass out to high redshifts. To understand the scatter that limits the used of HCN 1-0 as a proxy, it necessary to understand any environmental dependencies on the fidelity of this tracer. In this paper we present the first resolved comparison of HCN 1-0 with an independent tracer of dense gas mass in an extreme environment: the Galactic center (the central $\sim$ 300 pc of the Milky Way), a region with physical properties (high densities, temperatures, and turbulence) that are analogous to those in high-redshift galaxies \citep{KL13}. We compare the HCN 1-0 intensity in individual giant molecular clouds with the molecular gas mass from Herschel column density maps, and report on variations in the ratio of these two quantities: the dense gas conversion factor. We also compare the behavior of HCN 1-0 to that of other 3 mm tracers, including \hcniso\, 1-0, \hco\, 1-0 , HNCO 4$_{04}$-3$_{03}$, and \cyano\, 10-9. Finally, we discuss the implications of this study for the interpretation of past and future studies using HCN 1-0 as a tracer of dense gas in the extreme environments of luminous infrared galaxies and high-redshift systems. 

\section{Data and Methods}
\label{obs}

\subsection{Molecular line data}
\begin{figure}
\includegraphics[scale=0.85]{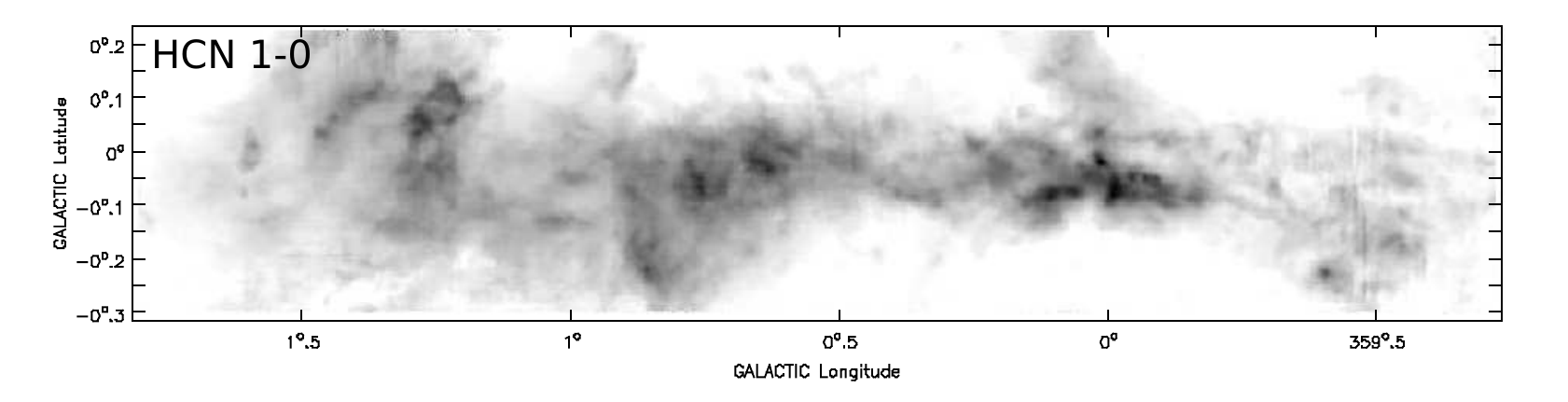}
\includegraphics[scale=0.85]{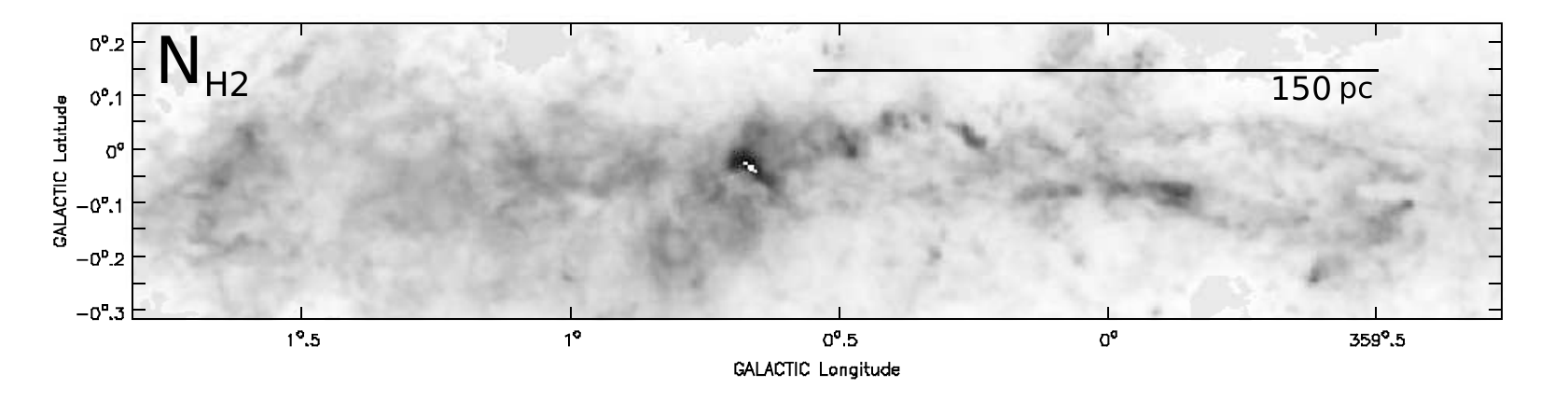}
\caption{A map of the integrated line luminosity of HCN 1-0 in uncorrected antenna temperature T$_A$* ({\bfseries Top}) is compared to a map of the total column density ({\bfseries Bottom}, Battersby et al. in prep.)}
\label{column}
\end{figure}

The first source of data for this analysis is a publicly-available 3 mm survey of the Galactic center with the 12 m Mopra telescope \citep{Jones12}. These observations covered frequencies from 85.3 to 93.3 GHz, which in addition to HCN 1-0 includes observations of \ctwo\, 1-0, \cthree\, 2$_{12}$-1$_{01}$, \methcy\, 5$_K$-4$_K$, \methacet\, 5$_K$-4$_K$, \cyano\, 10-9, \hoco\, 4$_{04}$-3$_{03}$, HNCO 4$_{04}$-3$_{03}$, \hco\, 1-0, \diaz\, 1-0, HNC 1-0, \hnciso\, 1-0, \hcoiso\, 1-0, \hcniso\, 1-0, \csiso\, 2-1, SO 2$_2$-1$_1$, and SiO 2-1. The velocity resolution of these data is $\sim$2 km/s. The typical angular resolution of these data is 36$''$ which, at an assumed distance of 8 kpc to the Galactic center \citep{Ghez08,Boehle16}, corresponds to a spatial resolution of 1.4 pc. The total extent of the mapped region spans 70 pc in Galactic latitude, and 370 pc in Galactic longitude. For the purposes of this analysis, we integrate the data cubes for each line over a velocity range [-230 \kms,+230\kms] sufficient to include all CMZ gas, discarding emission below a conservative 5-sigma cut off ($\sim$0.34 K \kms) for the lines with more prominent spurious features (HCN, \hco, HNC, \hcniso, HNCO, and \cyano; see discussion in Jones et al.), and a smaller cut off of 3-sigma (0.2 K \kms) for the remaining, primarily weaker lines. This is sufficient to eliminate the effects of artifacts such as anomalous baseline shapes in all but the weakest lines (e.g., \ctwo, \cthree). Following \cite{Jones12}, we do not correct for the main beam efficiency of Mopra, which for the extended emission in these maps will be $\sim$0.65 \citep[as measured at 86 GHz by][]{Ladd05} and variable with frequency. Line luminosities in this paper are then given in intrinsic units of uncorrected antenna temperature (T$_A$*) rather than main beam temperature.  The resulting integrated intensity map of HCN 1-0 is shown in Figure \ref{column}.

\subsection{Dust column density data}
The second source of data is an H$_2$ column density map (Figure \ref{column}, Battersby et al. in prep), which is derived from measurements of the submillimeter dust emission made with Herschel Hi-Gal in the Galactic center \citep{Molinari11}. At the wavelengths observed with Herschel, there is a significant contribution from Galactic cirrus emission. In \cite{Battersby11}, the background was defined (and later subtracted) through 16 iterations, however, in the current version, the process iterates dynamically until the fluctuations in the final cutoff stabilize, i.e. the standard deviation over the last 3 iterations is less than 1 MJy/Sr.  This variable iteration improves the overall contiguity of the method over different regions in the Galaxy.  While the background/foreground contribution is uncertain, this new variable iteration method allows a great degree of flexibility in background removal based on Galactic environment.  We do not expect that all of the Herschel emission at this longitude range is from the Galactic Center, but, in this region, the contribution from dense sources is much greater than the contribution from Galactic cirrus, so we expect uncertainties in the cirrus removal to be minor.  Moreover, any systematic differences in the Galactic cirrus toward the Galactic Center from our model would affect all of our targets, so the relative column densities would remain unchanged.

The resolution of the resulting column density map is 36", identical to the resolution of the Mopra data. Note that data is lacking from the central pixels of the brightest submillimeter source, Sgr B2, because the emission in these pixels is saturated. As a result, these pixels are masked, and the total mass of Sgr B2 will be underestimated. The method assumes a gas to dust ratio of 100 and dust with a $\beta$ of 1.75, as described in more detail in \cite{Battersby11}.  The median of the formal errors in the column density is $3\times10^{21}$ cm $^{-2}$, however, the error is a strong function of parameter value, with typical percentage errors of about 70\%. Uncertainties
in the background/foreground determination are not known a priori, however since the background/foreground model is subtracted from the entire region, any changes would likely affect the absolute, not relative column densities.

\subsection{Analysis Methods}
As the molecular line and column density maps have identical resolution, the first step is to identically pixelize the maps. Ratio maps are then constructed, representing the correspondence between the intensity of each line and the dense gas mass, which is proportional to the H$_2$ column density at each position. Only those pixels in all maps that have an H$_2$ column density above a chosen threshold (N(H$_2$) = 7$\times10^{22}$ cm $^{-2}$) are retained in the main analysis; this threshold is selected in order to isolate the emission from the most massive cloud cores, and the resulting thresholded ratio maps are shown in Figure \ref{ratio}. 

We employ two methods for comparing the variations in the ratio of HCN 1-0 intensity and N(H$_2$): first, we plot the base-10 logarithm of the total HCN 1-0 intensity and the base-10 logarithm of the total mass of H$_2$ in Figure \ref{clouds} for a sample of 13 giant molecular cloud cores: GCM1.6-0.03, Sgr D, GCM0.83-0.18 (a ring of gas and dust to the southwest of Sgr B2), Sgr B2 -- both its dense core (R$\sim$15 pc) and its `halo' (R$\sim$50 pc), GCM0.50+0.00 (the easternmost cloud in the dust ridge, near Sgr B1), the remaining dust ridge clouds, GCM0.25+0.01 (the `Brick'), GCM0.11-0.08 (together with two other neighboring clouds to the east of the 50 \kms\, cloud), GCM0.07+0.04 (A cloud on the edge of the Arched filaments), GCM-0.02-0.07 (the 50 \kms\, cloud), GCM-0.13-0.08 (the 20 \kms\, cloud), and Sgr C. The locations of these clouds are shown in the first subplot of Figure \ref{ratio}. In choosing to focus this analysis on the cores of these clouds above a given column density, the goal is to ensure that HCN 1-0 intensity is primarily being compared just to the mass of dense gas in the Galactic center. While the column density should in principle be sensitive to the mass of gas at any density, the cores of these clouds are believed to have average densities $>10^4$ \cm\, \citep{Gusten83,Zylka92,Serabyn92,Longmore12a,Longmore13b}.

Note that the line intensity and the column density are integrated over a region in the cloud above a given column density threshold, so the masses given here are equivalent to the mass of a subset of the cloud core, and are not equivalent to the total cloud masses. The column density thresholds used are 2$\times10^{23}$ cm$^{-2}$ for the core of Sgr B2 and GCM0.50+0.00, in order to separate them from the surrounding cloud `halo', and 7$\times10^{22}$ cm$^{-2}$ for all other clouds. The `halo' of Sgr B2 is defined by gas above this column density threshold over the entire contiguous Sgr B2 complex, but excluding the two cores (Sgr B2 and GCM0.50+0.00) where the column density is larger than 1.5$\times10^{23}$ cm$^{-2}$. This threshold is chosen so that the mass of the Sgr B2 core and halo are roughly equivalent. Total masses are then calculated from the integrated H$_2$ column density according to the formula:

\begin{equation} M_{\mathrm H2} = \mu_{\mathrm H2} \hspace{0.1cm} m_{\mathrm H} \int N_{\mathrm H2} \hspace{0.1cm} dA
\label{eq1}
\end{equation}

where $\mu_{\mathrm H2}$ is the mean molecular weight of the gas, which we take to be 2.8, assuming
abundances of 71\% H, 27\% He, and 2\% metals \citep[e.g.,][]{Kauffmann08}, $m_{\mathrm H}$ is the mass of a Hydrogen atom, and $\int N_{\mathrm H2} \hspace{0.1cm} dA$ is the column density of H$_2$ (in cm$^{-2}$ integrated over the area of the cloud (in cm$^2$). Note that we are assuming in this calculation that all of the observed clouds are at the same distance, 8 kpc.

Second, we analyze the pixel-by-pixel variation of the HCN 1-0 intensity (shown in Figure \ref{pixels}) in which this quantity is plotted against N(H$_2$). To aid in the interpretation of these plots we additionally colorize points if they correspond to one of the thirteen selected clouds. Note that pixels in the core of Sgr B2 for which the column density value is saturated are not included in this plot. 

\section{Results}	
\label{res}

Comparing log$_{10}$ (HCN 1-0) with log$_{10}$ (M$_{H2}$) in Figure \ref{clouds} we find a largely linear relationship over 2 orders of magnitude in mass. However, we also see a (maximal) scatter of 0.75 dex in the HCN 1-0 line intensity at a given mass. Some notable sources contributing to this scatter include the core of Sgr B2, GCM-0.02-0.07, and GCM0.11-0.08. The core of Sgr B2 appears underluminous, having an integrated HCN 1-0 line intensity equivalent to that of clouds up to five times less massive. It also has five times weaker HCN 1-0 than the halo of the Sgr B2 cloud which contains a roughly equivalent mass of dense gas. Additionally, GCM-0.02-0.07 and GCM0.11-0.08 have integrated HCN 1-0 line intensities 3 times brighter than other clouds of comparable mass. 

Examining other 3 mm tracers, we find that HNCO  4$_{04}$-3$_{03}$ appears to have an even tighter linear relationship than HCN 1-0 (maximal scatter of only 0.5 dex ).  In contrast, \hco\, 1-0  exhibits at most nearly 1 dex of scatter. Interestingly, the $^{13}$C isotopologues of HCN exhibits much more scatter than the main $^{12}$C isotopologue: we see a maximum of around 1.1 dex of scatter in the line intensity at a given mass (which may be an underestimate of the true scatter, as a number of the more massive clouds detected in H$^{12}$CN are not detected in \hcniso). We also note that GCM-0.02-0.7 is overluminous in the majority of tracers compared to other clouds having a similar mass. Some of the largest variation in line intensity at a given cloud mass is seen in \cyano\, and \methcy, with both having  a maximum of 1.2-1.4 dex of scatter in these plots. 

In order to better quantify how well a linear fit describes the relationship between log$_{10}$( integrated intensity ) of each molecule and log$_{10}$( mass ), we determine the linear correlation coefficient (r) and the two-sided p-value for a hypothesis test whose null hypothesis is that the slope is zero. The r- and p-values for each molecule are given in Table \ref{stats}. The strongest linear correlation is seen for HNCO (r = 0.95), with HNC, HCN, HCO+, SiO and N2H+ being similarly strongly correlated (r$\sim$0.82-0.88). The remaining molecules having a linear correlation coefficient r $>$ 0.5 also have p $>$ 0.003 and cannot be distinguished with a 3 sigma certainty from the null hypothesis.  Of these, \hcniso, \hcoiso, \csiso, and \cthree have p-values greater than 0.1 (so the null hypothesis can only be ruled out with a 1 sigma certainty). For many of these lines, this large scatter may in part be due to the smaller number of clouds that are detected in these lines.  \methacet, \hoco, and SO are not detected in a sufficient number of clouds for the correlation to be examined. These results suggest that HNCO  4$_{04}$-3$_{03}$ (despite not being the fundamental rotational transition of that molecule) could be a comparable if not better tracer of mass than HCN 1-0 in the Galactic center environment. 

We also look at the pixel-by-pixel distribution of the HCN 1-0 intensity versus column density (shown on a log-log plot in Figure \ref{pixels}) in these Galactic center clouds.We find that pixels in individual clouds also appear to show a linear relationship between the logarithm of the intensity of HCN 1-0 and the logarithm of the corresponding molecular gas mass (where the plotted column density for a uniform-sized pixel is proportional to this mass). However individual clouds are offset by 0.7-0.8 dex in their HCN 1-0 intensity. We also see that the scatter in HCN 1-0 intensity for an individual cloud is not constant: it varies from $\sim$ 0.1 dex in GCM-0.02-0.07 up to 0.5 dex in GCM0.25+0.01 and higher column-density regions of GCM1.6-0.03. This indicates that variations in the ratio of HCN 1-0 intensity to gas mass are present not just on the scales of entire Galactic center clouds ($\sim$ 10 pc) but also on scales down to a few parsecs. 

While we have focused our analysis on the cores of well-known clouds which should be representative of the densest gas in the Galactic center, only 32\% of the total gas mass in the Galactic center as traced by the column density map lies above the chosen column density threshold of N(H$_2$) = 7$\times10^{22}$ cm $^{-2}$. We therefore also investigate the behavior of the more diffuse or extended component in HCN 1-0. The ratio of HCN 1-0 intensity to the column density above a threshold of 1$\times10^{22}$ cm $^{-2}$ is shown in Figure \ref{IR}. The ratio varies by more than two orders of magnitude: it reaches a minimum of $\sim$0.5 (the units on this ratio are K \kms\, per 10$^{22}$ cm$^{-2}$) in Sgr B2,  and maxima of $\sim$ 60 are reached in the gas infrared bubble \citep{RF01b} and north of Sgr D. These extended regions have ratios more than 3$\times$ greater than those found in any of the cloud cores. Overall, a  HCN 1-0 is slightly brighter in the lower column density gas: 80\% of the HCN 1-0 emission comes from gas with column density less than 7$\times10^{22}$ cm $^{-2}$, accounting for 68\% of the total mass, while only 20\% of the HCN 1-0 comes from gas with column density above 7$\times10^{22}$ cm $^{-2}$, accounting for 32\% of the total gas mass.

\section{Discussion}
\label{dis}

Although the brightness of HCN 1-0 appears to correlate well with the gas mass over several orders of magnitude in Galactic center clouds, we also see a few individual clouds that are outliers from this relationship. We thus focus on these clouds in an attempt to characterize the origin of the scatter in the relationship between HCN 1-0 and gas mass in this environment. 

\subsection{Sgr B2: Underluminous HCN 1-0 in the most massive Galactic center cloud}
One of the most surprising results of this analysis is the relative weakness of 3 mm line emission from the core of Sgr B2, the most massive molecular cloud in the Galactic center.   

Examining the spectra of all detected transitions toward the core of Sgr B2 (Figure \ref{B2}), we note that the shape of many transitions have a two-peaked structure, with a pronounced dip at the central velocity of the bulk of the gas in Sgr B2 \citep[$\sim$ 65 \kms, as traditionally determined from observations of absorption toward the \hii regions in Sgr B2 using tracers such as NH$_3$, e.g.][]{Wilson82}. While there is a second velocity component also observed in absorption toward Sgr B2 at $\sim$ 80 \kms \citep{Win79}, the observed two-peak structure is not consistent with the detection of these two kinematic components. Instead, the symmetry of the two peaks around a central velocity at which more optically-thin tracers (e.g., less abundant and more complex molecules) have a single peak is a clear signature of self absorption. In self absorption, cooler or lower-excitation gas is present along the line of sight and at the same velocity as warmer or higher-excitation gas. The foreground column of cooler molecules then absorbs the emission signature from the more excited background molecules.  As would be expected for self absorption, the transitions which are most affected are the lowest-J transitions: All species whose 1-0 transition is observed (HCN,\hco,HNC, including the $^{13}$C isotopologues of these species, \ctwo\, and \diaz) exhibit this dip. The dip is also seen in the 2-1 transitions of $^{13}$CS, SiO, and \cthree\, though not in the 2$_2$-1$_1$ line of SO. However, it is not present in \hoco, \methcy, \methacet, \cyano, or HNCO, all of which are observed in transitions having J$_{up}\ge$4. We thus favor self absorption as the cause of the apparently weak emission of HCN 1-0 in Sgr B2. Less prominent self-absorption is also seen in the spectra of HCN 1-0 toward GCM1.6-0.03, GCM-0.13-0.08, and Sgr C, though none of these sources also have self-absorption in \hcniso. 

Self-absorption in higher-J lines of HCN and \hco\, has also been observed in the Galactic center `Circumnuclear Disk' surrounding the supermassive black hole, however here only the $^{12}$C lines are self-absorbed \citep{Mills13}. The observed self absorption of even the $^{13}$C isotopologues toward the Sgr B2 core then likely requires a higher column of self-absorbing gas toward this source. Given that the self-absorption happens at the same velocity as the peak emission in the Sgr B2 core, it is likely that this is a fairly localized effect (i.e., this is not a resulting of intervening, foreground gas in the Galactic disk). We note that independent of the observed self-absorption there is also likely absorption against the embedded HII regions in Sgr B2 -- however, as these HII regions are more centrally concentrated than the self-absorption signature, and there are molecules for which dip is not seen, we do not believe the observed signature can be due to absorption against HII regions alone. 

Higher-excitation tracers of more common molecules are also likely to be less affected by self-absorption. For example, in observations of the Galactic center circumnuclear disk while self-absorption is seen in the J = 3-2 and 4-3 lines of HCN, self absorption is no longer present in the 8-7 line \citep{Mills13}. This could make diatomic tracers of dense gas such as CS a more optimal tracer in a given band, as they will have higher-J transitions than HCN, HNC, or \hco at similar frequencies. In the observations analysed here, the 2-1 transition of \csiso\, can be seen in Figure \ref{B2} to have less self absorption than the 1-0 transitions of \hcniso, \hnciso, and \hcoiso. However, care should still be taken in avoiding tracers subject to self-absorption to also avoid tracers such as SiO that are known to be enhanced by shocks.  

The fact that the core of Sgr B2, the most massive and dense cloud in the Galactic center is underluminous in HCN 1-0 has important implications, as it is often taken to be a template for conditions that are widespread in more extreme starbursts or high-redshift systems. This result indicates that caution should be taken when interpreting variations of the ratio of HCN 1-0 to far-infrared luminosity in extreme systems, as they could instead be due in part to HCN 1-0 underestimating the total dense gas mass. We discuss this further in Section \ref{implications}.

\subsection{Overluminous HCN in Galactic center clouds}

While self-absorption appears to be the dominant effect contributing to the weakness of integrated HCN 1-0 emission from Sgr B2, we also see variation in the relationship between the HCN 1-0 line strength and the gas mass for other Galactic center clouds in our survey. Outside of the Sgr B2 core, three dense clouds: GCM-0.02-0.07, GCM0.07+0.04, and GCM0.11-0.08, have brighter HCN 1-0 emission by a factor of $\sim$3 than other clouds of comparable mass, including clouds that show no signature of self-absorption (e.g., GCM0.25+0.01). 

While the observed overluminosity of HCN 1-0 in several Galactic center clouds is small (on the order of a factor of 3), determining the mechanism that is responsible is important for assessing whether this effect might become more pronounced in other environments. Identifying the mechanism which is leading to enhanced HCN 1-0 can also give insight into additional diagnostics which should be performed in order to assess whether HCN 1-0 is accurately representing the total dense gas mass in a given environment. 

\subsubsection{Variation in HCN 1-0 optical depth}

We first examine whether the relative weakness of HCN 1-0 in individual clouds could be due to a high optical depth of this line. In addition to HCN 1-0, we also observed \hcniso, which should be $\sim$ 25 times less abundant than H$^{12}$CN \citep{WR94, Wilson99,Riq10}. We find that the ratios of HCN 1-0 and \hcniso\, 1-0 do not show clouds with relatively weaker HCN 1-0 to be preferentially optically thick. Thus, optical depth variations are not a viable explanation for the observed variations in HCN 1-0 intensity as a function of cloud mass. In fact, the clouds with relatively bright HCN 1-0 for their mass are some of the most optically thick: GCM-0.02-0.07, which has some of the brightest HCN 1-0 emission is 10 times brighter in the optically-thin \hcniso\,1-0 line than the comparably-massive GCM0.25+0.01. This indicates that the optical thickness of HCN 1-0 in a number of clouds is actually masking a greater underlying variation in the intensity of HCN lines. We suggest then that although GCM-0.13-0.08 does not appear overluminous in HCN 1-0 for its mass, this is only due to the optical thickness of the HCN 1-0 line in this source, which is comparable to that in GCM-0.02-0.07 (HCN 1-0 / \hcniso\, $\sim$ 5-6). This would then mean that all of the clouds projected to lie within $\sim$ 30 pc of the supermassive black hole Sgr A* have enhanced HCN 1-0 emission for their mass. 

\subsubsection{Variation in excitation conditions}

We next ask whether this difference could be due to varying excitation conditions for HCN 1-0. Other studies have shown that there is no significant difference in the kinetic temperature \citep{Huttem93b,Ao13,MM13,Ginsburg16} or line width \citep{Jones12} between the clouds in our study with relatively weak and bright HCN 1-0. Our RADEX analysis then indicates that for a constant column density, temperature, and linewidth, variations in volume density from n=$10^5-10^7$ should change the line brightness by less than a factor of 2. In this case, the observed overluminosity can only be explained if the three bright clouds are the only clouds (apart from the Sgr B2 core) with n $> 10^{4.5}$ \cm. Independent studies of cloud densities \citep{Serabyn92,Longmore13b,Gusten83,Zylka92} are inconsistent with this scenario, and in fact even higher ratios of HCN 1-0 to N(H$_2$) are seen in the more extended gas which is likely to be at yet lower densities. We note that there could still be excitation differences due to radiative excitation or masing, however for either of these to be applicable, HCN 1-0 should be well correlated with IR luminosity.

Our RADEX estimates could however be changed if the filling factor of HCN 1-0 (here assumed to be 1 in all of the observed clouds) varies significantly from cloud to cloud. In the case of a smaller filling fraction, the column density of HCN 1-0 would be higher for the same observed line brightness. However, observations of several Galactic center clouds in lines of NH$_3$, which should trace a similar dense gas component as HCN 1-0, show that for all clouds observed so far, the bulk of the gas emission is in an extended component \citep[The fraction of emission on scales larger than a few tens of arcseconds that is resolved out by interferometric observations is 75-95\%][]{Armstrong85,MM13}.This is consistent with the assumed filling fraction of 1. As the fraction of emission resolved out by interferometric observations has not been measured for all clouds in our study, and this measurement has not been done for HCN, it is possible that this effect could contribute to the observed variation in the relative brightness of HCN 1-0. However, we do not believe that it is likely that this is the dominant source of the observed variation. 

\subsubsection{Variation in conditions in the local environment}

We then examine whether the variations in the HCN 1-0 intensity could be consistent with a dependency of this quantity on the environment. It has been suggested that overluminous HCN could be due to X-ray chemistry \citep{LD96,Meijerink07,Harada13}. While several regions of enhanced emission are associated with X-ray sources \citep[GCM-0.02-0.07 borders a supernova remnant, and clouds in the GCM0.11-0.08 complex are observed to have a propagating X-ray light echo][]{Ponti10} that may also be related to enhanced SiO in these regions \citep{JMP00}, there is no X-ray emission associated with the bright extended HCN 1-0 emission north of Sgr D \citep{Ponti15}. We also investigate whether the enhanced HCN 1-0 is correlated with infrared emission. Clouds that are actively forming stars have both weak and bright HCN 1-0, so an internal IR radiation field is likely not an important factor. The three clouds with brightest HCN 1-0 (G0.07+0.04, GCM0.11-0.08, and GCM-0.02-0.07) are projected against the infrared bubble. While this might suggest a connection between HCN 1-0 and the presence of a strong external IR field, the region north of Sgr D has extremely elevated HCN 1-0 in Figure \ref{IR} without significant IR emission. We also note that although Sgr B2 has both a strong internal infrared field from embedded formation and a recent X-ray light echo observed to be propagating through the cloud \citep{Terrier10}, self absorption prevents us from determining whether HCN 1-0 is enhanced in the Sgr B2 core.  Based on these data we then find that the HCN 1-0 intensity does not appear to depend upon the local IR radiation field or nearby X-ray sources. However, given that all of the clouds with enhanced HCN 1-0 emission are located near each other, and projected within 30 pc of the supermassive black hole, we cannot rule out an importance of other environmental factors. 

\subsubsection{Variation in HCN abundance}

Given that the observed differences in HCN 1-0 intensity for a given cloud mass do not appear likely to be a function of optical depth or excitation variations (including radiative excitation or masering, as the HCN 1-0 emission is not uniformly correlated with strong IR emission), we suggest that they are a function of varying HCN abundance. Although the scale of variations in the brightness of HCN 1-0 for a given cloud mass (apart from Sgr B2 and regions affected by self-absorption) is only a factor of 3, we will assume that the true scale of variations in the abundance of HCN in Galactic center clouds is at least an order of magnitude, as is measured in the optically-thin \hcniso\, line.

One possibility for explaining overluminous emission from HCN 1-0 is that HCN  is preferentially enhanced in some clouds and regions of the Galactic center due to elevated shock activity. HCN abundances have been observed to be enhanced by 1-2 orders of magnitude in shocked environments ranging from protostellar outflows to extragalactic outflows \citep{Jorg04,Tafalla10,Aalto15a}. GCM-0.02-0.07 borders a supernova remnant and shows multiple signs that it is experiencing shocks from this interactions \citep{Sjou10,YZ96}, and while the GCM0.11-0.08 complex is not suggested to be interacting with a supernova remnant or to have undergone a cloud collision, it could be possible to explain enhanced shock activity in these clouds due to interaction with the expanding IR bubble, or as a result of their recent pericenter passage in the model of \cite{Kruijssen15}. Importantly, the enhanced HCN 1-0 emission north of Sgr D is also coincident with SiO emission, which is often indicative of shock chemistry \citep{JMP97}. The other large area of extended enhanced HCN 1-0 emission visible in Figure \ref{IR} is largely centered on the nucleus. A direct link to shock activity is not immediately apparent, although it is possible to hypothesize several shock-related scenarios: it could be related to shocks driven by stellar winds, given that it correlates will with the IR emission tracing the influence of centrally-concentrated young stars and clusters. Or, it could be linked to shocks from a dissipation of turbulent energy fed by the change in shear as gas migrates inward from the 100 pc ring \citep{Molinari11, Kruijssen15} toward the central supermassive black hole \citep{Wilson82}. 

As Sgr B2 has been suggested to have undergone a cloud-cloud collision, this cloud might also be expected to host enhanced shock activity-- and indeed, it is possible that in the Sgr B2-halo, which is largely free of self-absorption, that the HCN 1-0 line is overbright (SiO and HNCO, another shock tracer, both appear to be strong and perhaps overluminous in this region as well).  Additional work is needed to confirm this hypothesis: for example, shocks might be expected to correspond with regions of increased turbulence that could manifest as increased line widths. Enhanced emission from other shock tracers could also be searched for, such as the 36 GHz methanol maser, which has been recently observed to be widespread in the Galactic center \citep{YZ13,Mills15}. 

\subsubsection{Variation in dust properties}

Finally, we also investigate whether the observed variations in the ratio of HCN 1-0 intensity to cloud mass could be a function of errors in the derivation of the cloud masses themselves. The results of this study (attributing these variations to differences in the intensity of HCN 1-0) are based upon the assumption that the submillimeter dust emission traced by Herschel is a high-fidelity tracer of the total gas column density (and thus mass) in the Galactic center. We cannot, at this time, rule out that variations in the gas to dust ratio (we assume 100 in this work) or dust emissivity (we assume $\beta$ of 1.75) could be partly responsible for some of the discrepancies we see.  However, we note that these values would need to vary on a cloud-to-cloud scale to explain our results, as the relative column densities of the clouds do not change if the global Galactic Center value of the gas to dust ratio is revised.

The assumption that the dust is a good tracer for the gas column density appears to be valid in less-extreme environments in our Galaxy \citep[e.g.][]{Battersby14}. Bulk dust properties have also been shown to be extremely similar in environments ranging from our Galaxy to nearby normal star forming galaxies and submillimeter galaxies \citep{Scoville16}. The strongest argument for the Galactic center dust continuum being a good tracer of mass in the Galactic center that dust-derived cloud masses of this study are consistent with cloud masses previously inferred using C$^{18}$O \citep{Dahmen}. For example, the approximate mass of Sgr B2 complex estimated from the dust continuum (a few $10^6$ \msun) is comparable to that determined using optically-thin measurements of C$^{18}$O \citep{Dahmen}. We thus do not believe that the use of dust emission to determine masses affects the main results of this paper, although the assumption that the dust to gas ratio does not significantly vary as a function of environment has yet to be carefully tested in the Galactic center.

\subsection{Implications of a varying HCN 1-0 to dense gas mass conversion factor}
\label{implications}

We have found that the logarithm of the HCN 1-0 intensity is linearly correlated with the logarithm of the dense molecular gas mass over two orders of magnitude in the gas mass in Galactic center molecular clouds. However individual clouds in this environment show deviations from this relationship that contribute to around 0.75 dex of scatter. The two primary sources of scatter that we have identified are self-absorption (which reduces the observed HCN 1-0 intensity of Sgr B2) and variations in HCN abundance (possibly due to shock enhancements), which appear to increase the brightness of HCN 1-0 in GCM-0.02-0.07 and other clouds projected to lie within the central 30 pc of the Galactic center. 
Here, we discuss the implications of our observations for interpreting the Gao-Solomon relation and for generally deriving accurate total gas masses from 3 mm tracers.

\subsubsection{The interpretation of the Gao-Solomon relation}

The scatter that we measure in the relationship between HCN 1-0 intensity and dense gas mass for 13 Galactic center clouds ($\sim$ 0.75 dex) is actually less than the scatter measured by \cite{Wu10} in their Figure 43 ($\sim$ 1.5 dex), for a sample of a few dozen Galactic clumps. There are several differences between our sample and that of \cite{Wu10}. In our sample, the clouds have typical radii of 3-5 pc and typical masses of 10$^5$ \msun, with masses spanning two orders of magnitude (from a few $10^4$ - $10^6$ \msun). The clump sample of \cite{Wu10} in contrast has typical radii of 1 pc and typical masses around 1000 \msun (spanning 4 orders of magnitude from $\sim$10 to $10^5$ \msun). The masses measured by \cite{Wu10} are also virial masses, measured from the linewidth of C$^{34}$S 5-4. Studies of the relationship between the virial mass and the actual mass (the virial parameter) over a range of environments and cloud masses show that there is significant scatter in this relationship \citep[e.g.,][]{KPG13,Battersby14}. In particular, \cite{KPG13} find $\sim$2 orders of magnitude of scatter in the virial parameter for masses comparable to those in the sample of \cite{Wu10}, which may contribute to the larger scatter that they observe in their relation between HCN 1-0 and dense gas mass. We conclude then that the results from these samples are not directly comparable, but we note that the scatter we have identified for Galactic center clouds is not unprecedented, and may indeed be less than that already observed on smaller scales. 

Identifying the origin of this scatter is important for the related Gao-Solomon relation, which shows that the HCN 1-0 integrated intensity is positively correlated with the far infrared luminosity over 10 orders of magnitude in IR brightness \citep{GS04b,Wu05}. This relation is typically interpreted as a relationship between the total amount of (dense) molecular gas and star formation, where HCN 1-0 is a proxy for the amount of this gas and the IR luminosity is a proxy for the amount of star formation. Assuming a fundamental relationship between the amount of dense gas and the star formation rate exists, then scatter in the relationship between the HCN 1-0 intensity and the actual amount of dense gas would contribute to the observed scatter in the Gao-Solomon relationship. Our identification of HCN abundance variations and self absorption as sources of this scatter could help explain the increased scatter and systematic deviations seen when testing the Gao Solomon relationship for larger samples of more extreme sources, including ULIRGs \citep{Privon15} and AGN \citep{Davies,Imanishi09,Kohno03}. 

\cite{Wu05} interpret the Gao-Solomon relationship as indicating that individual, sub-parsec scale dense cores are the fundamental unit of dense gas mass associated with star formation, and that the Gao Solomon relation on larger scales is essentially counting the number of these structures. However, our observations that only 20\% of the HCN emission in the Galactic center comes from the cores of the Giant molecular clouds (above a column density threshold of N(H$_2$) = 7$\times10^{22}$ cm $^{-2}$) challenge this interpretation. Instead, the bulk of the molecular gas traced by HCN 1-0 in this environment lies outside of these dense clouds, the structures in which all observed Galactic center star formation is occurring. This indicates that in the environment of the Galactic center, the Gao-Solomon relation cannot be entirely a relationship built up from the star-forming properties of individual dense, high-mass clumps. This is consistent with a recent study of the clump-scale Gao-Solomon relation using the MALT90 survey data of the Galactic plane, which found that if the Galactic-scale Gao-Solomon relation were due just to individual star forming clumps, it would require far more high-mass star forming clumps than are actually observed. \citep{Stephens16}. Any correlation between HCN 1-0 and IR emission in this region must be explained in the context of the dominant extended molecular gas component that is not currently associated with the most massive star forming clouds, and the extended IR emission from past generations of star formation \citep[as indeed, the bulk of even the dense clouds are not actively star forming here, e.g.,][]{Lis01,Longmore13a}.

\subsubsection{Deriving accurate total gas masses from 3 mm tracers}
In the center of our Galaxy, scatter in the relationship between HCN 1-0 and total cloud mass would not contribute to a large inaccuracy of the mass inferred just using HCN 1-0 intensity. Although the Sgr B2 core has HCN 1-0 emission consistent with a cloud $\sim$5 times less massive, this missing luminosity would only result in missing $\lesssim10$\% of the total mass of dense gas in the Galactic center. Some of this could also be balanced out by the ``extra mass" inferred from overbright HCN 1-0 in other clouds. However, in more globally-extreme environments than the Galactic center, what we observe as scatter could lead to systematic errors when interpreting HCN 1-0 as proportional to the total mass of dense gas. 

Certain environments in particular should be more prone to the self-absorption in HCN 1-0 observed toward the Sgr B2 core. In clouds having an excitation gradient that increases radially toward the center, one might expect to see self-absorption in the lower J levels of a given common gas tracer. Such excitation gradients would be typical of clouds with embedded active star formation (as in Sgr B2), or embedded AGN, as is has recently been noted  in a number of compact sources of nuclear emission \citep{Aalto15b}. Self-absorption could also occur in cases where there are strong global signatures of outflow or infall. Self absorption also might be expected to depend on the relative orientation of a warm background source and cool foreground emission, but as it requires the foreground and background gas to be at the same velocity, it should not be a general feature of galaxy disks observed edge-on (for example, we note that self-absorption is only seen for a small number of the surveyed Galactic center clouds, though all are observed through edge-on disk of our own Galaxy). 

If the enhanced HCN abundance is linked to nuclear activity, as the concentration of clouds overluminous in the HCN 1-0 line projected within 30 pc of Sgr A* might suggest, then nuclear starbursts or AGN might also experience a similar effect. In this case, HCN could systematically overestimate the total dense gas mass present. We thus recommend that observations of HCN 1-0 in extreme environments should be checked for systematically weak emission due to self-absorption (e.g., via comparison with tracers like \cyano\, or \methcy) before seeking a physical explanation for a lack of dense gas such as variations in the gas depletion time \citep[e.g.,][]{GB12,Juneau09,Leroy15,Usero15}. Such studies will also be important for determining whether the Galactic center conditions responsible for variations in the HCN 1-0 to dense gas conversion factor are actually relevant for other galaxies, or whether deviations from the Gao-Solomon relation in other systems are due to entirely separate effects that are a result of environments that are not sampled in our own Galaxy. 

In general, care should be taken in using any single transition to represent total mass: although \cyano\, and \methcy\, are better tracers of the total mass of the Sgr B2 core because of their immunity from self-absorption in this source, the overall intensity of both tracers in other Galactic center clouds are extremely poorly-correlated with total mass \citep[perhaps related to observations that \cyano\, is enhanced in active nuclei][]{Aalt07b}. Our findings indicate that, in order to determine accurate gas masses, one should primarily use a tracer like HCN 1-0, and supplement it with a tracer like \cyano\, or \methcy\, to guard against underestimation of the dense gas mass due to the effects of self-absorption. Further study of the correspondence of HCN 1-0 with shocks in extreme extragalactic environments should also be made to determine whether bright tracers that are similarly well-correlated with mass like HNCO 4$_{04}$-3$_{03}$, HNC 1-0, \hco 1-0, or CS 2-1 are less affected and thus more suitable for determination of accurate gas masses in environments dominated by shocks. The especially good correspondence of HNCO with dense gas in this environment is worthy of further investigation. HNCO is suggested to behave like a shock tracer in some environments, based on its correlation with known shock tracers like SiO \citep[e.g.,][]{Zinchenko00,Meier05,Li13}. The observed correspondence may then simply be confirming that shocked gas conditions are ubiquitous in Galactic center gas \citep[as suggested by prior observations of SiO and CH$_3$OH][]{JMP97,YZ13} Interestingly, \cite{Henshaw16} also found HNCO 4$_{04}$-3$_{03}$ to be one of the best kinematic tracers of Galactic center clouds.

Finally, we also caution that some systematic deviations in molecular abundance may not be accounted for by employing multiple gas tracers, as we find that GCM-0.02-0.07 is overbright in nearly every 3 mm tracer (except perhaps HN$^{13}$C). This may be related to the unique chemistry or excitation in this cloud, which is by far the brightest source of C$^+$ emission in the Galactic center, likely due to its unique proximity to a nearby supernova remnant \citep{Tanaka11}. In such extreme cases, it may be that measurements of the mass using the submillimeter dust continuum \cite{Scoville16} provide the most accurate results. 

\section{Conclusions}

Below, we summarize the main findings of this paper:\\

\begin{itemize}

\item For our observed sample of 13 Galactic center clouds, log$_{10}$( HCN 1-0 ) is linearly correlated with the log$_{10}$ ( dense molecular gas mass ) over two orders of magnitude in the gas mass. However individual clouds in this environment deviate, in some cases systematically, from the relationship and contribute to around 0.75 dex of scatter. The two primary sources of scatter that we have identified are self-absorption and variations in HCN abundance. 

\item The core of Sgr B2 has an HCN 1-0 intensity equivalent to clouds having five times less mass. We find that the faintness of this line is a result of self-absorption in this source. Based on this, we caution against the interpretation of increased L$_{FIR}$/HCN 1-0 as indicative of a higher star formation efficiency or lower gas depletion time in extreme environments that might be dominated by sources like embedded starbursts or AGN having a similar excitation gradient, with more highly-excited gas surrounded by a lower-excitation envelope. The 3 mm lines of \cyano\, and \methcy\, do not suffer from this self absorption in Sgr B2, and so may be used to identify regions where HCN 1-0 is self-absorbed. 

\item Several cloud cores, all projected to lie within the central 30 parsecs of the Galactic center, have $\sim$3 times brighter HCN 1-0 than other clouds of similar mass. We find that this is likely a result of enhanced HCN abundance, and suggest that this could be a result of enhanced shock activity. 

\item We also find several contiguous regions of more extended gas emission that have 3-10 times more luminous HCN 1-0 per unit N(H$_2$) than in any of the cloud cores studied here. Overall HCN 1-0 is brighter in this lower column density gas: 80\% of the HCN 1-0 emission comes from gas with column density less than $7\times10^{22}$ cm$^{-2}$, accounting for 68\% of the total mass, while only 20\% of the HCN 1-0 comes from gas with column density above $7\times10^{22}$ cm$^{-2}$, accounting for 32\% of the total gas mass.

\item We examine other 3mm tracers and find that HNCO 4$_{04}$-3$_{03}$ shows a tighter correlation with cloud mass, suggesting that HNCO 4$_{04}$-3$_{03}$ (despite
not being the fundamental rotational transition of that molecule) could be a comparable if not better tracer of mass than HCN 1-0 in the Galactic center environment. HNC 1-0 and \hco\, 1-0 are less well-correlated than HCN 1-0. \cyano\, and \methcy\, show a poor correlation between their brightness and total mass, and should not be used to replace HCN 1-0, despite being free of self-absorption effects. 

\end{itemize}

\section*{Acknowledgements}
We thank the anonymous referee for detailed and insightful comments that substantially improved the presentation of the results in this paper. The authors also thank Jonathan Henshaw, Steven Longmore, David Meier, Susanne Aalto, and George Privon for useful discussion and suggestions that significantly improved the paper. This material is based upon work supported by the National Science Foundation under Award No. 1602583.

\bibliographystyle{hapj}
\bibliography{HCN}

\clearpage

\clearpage

\begin{figure}
\includegraphics[scale=0.85]{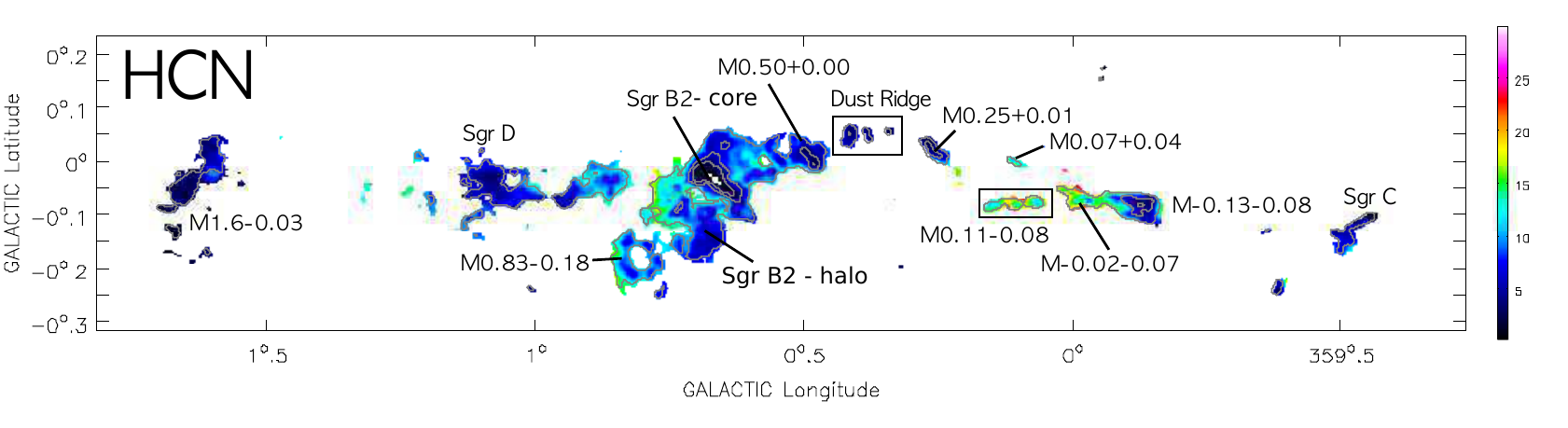}
\includegraphics[scale=0.85]{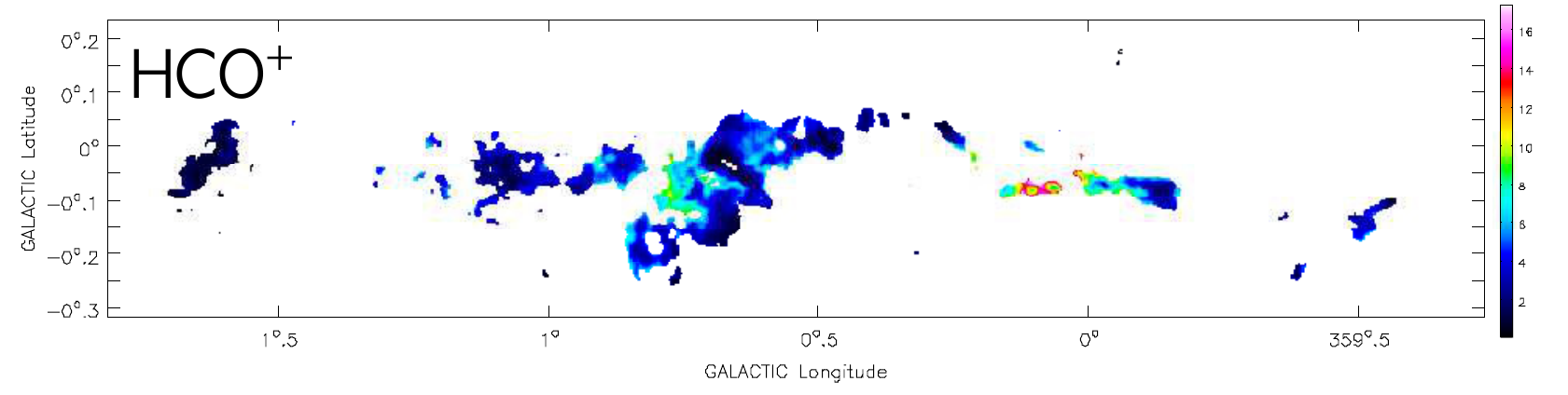}
\includegraphics[scale=0.85]{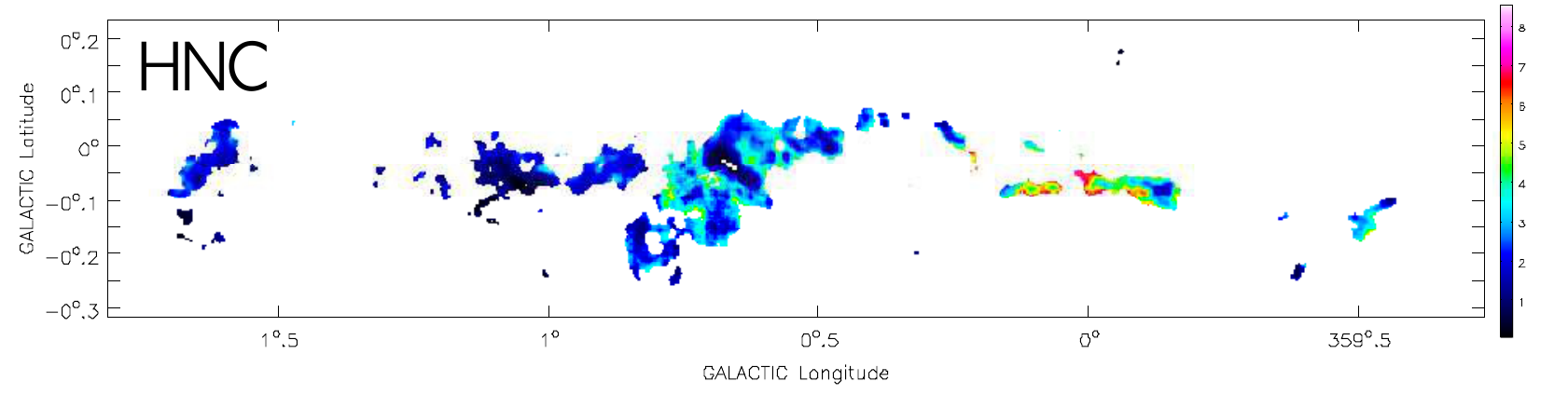}
\includegraphics[scale=0.85]{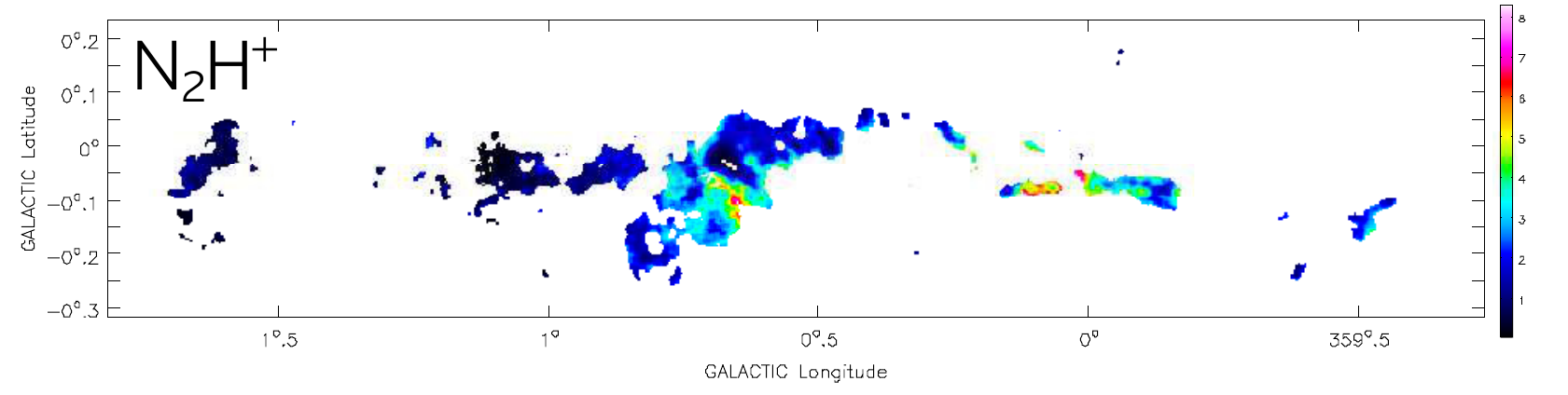}
\includegraphics[scale=0.85]{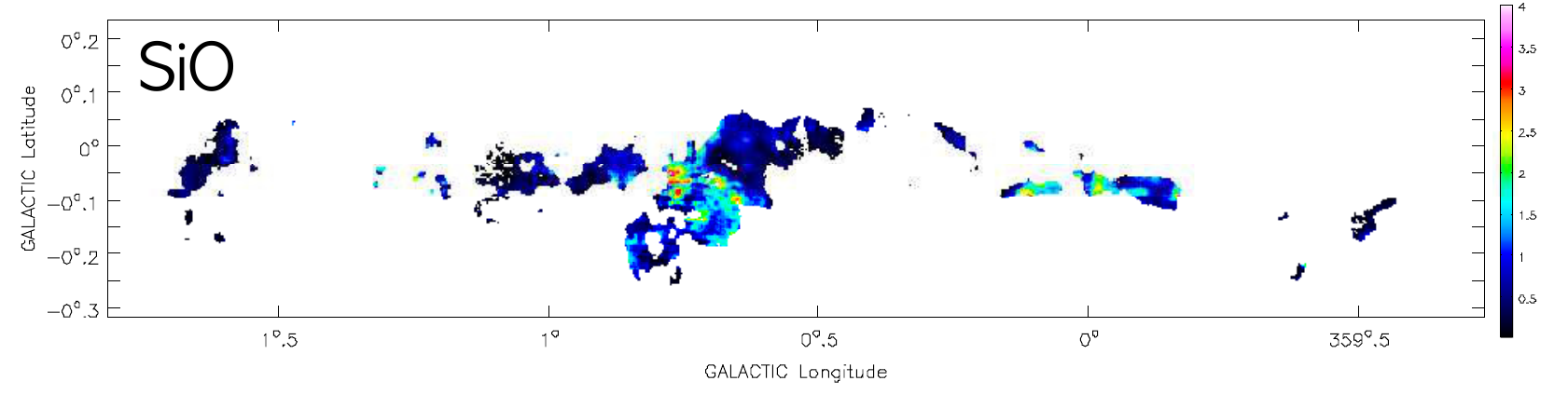}
\includegraphics[scale=0.85]{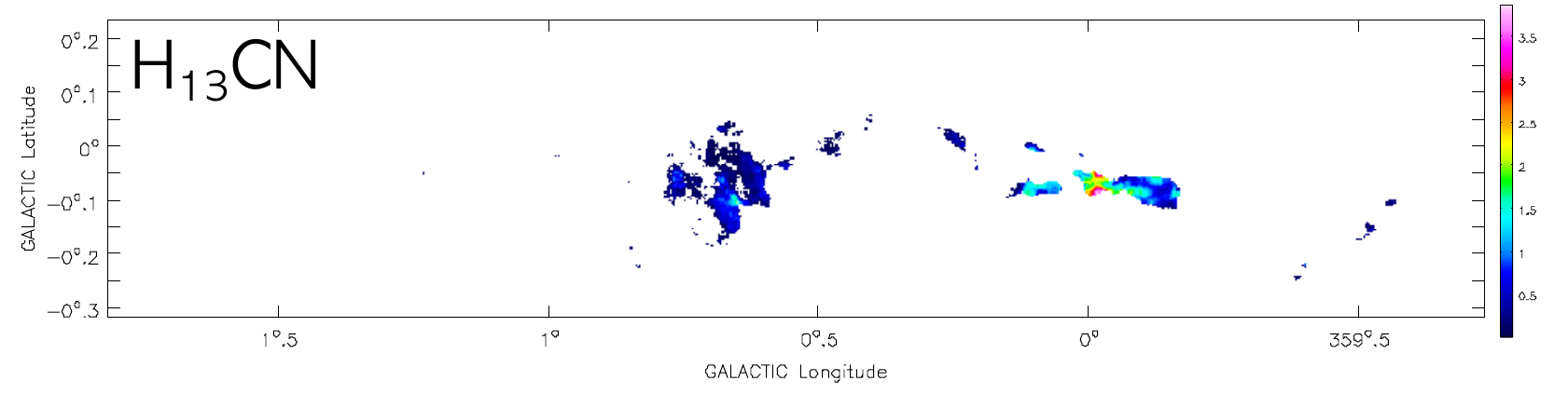}
\caption{Maps of the molecular line intensity to dense gas conversion factor. The units of the ratio shown are line intensity in K km/s over the column density in units of 10$^{22}$ cm$^{-2}$. The individual clouds studied in this analysis are identified in the first subfigure (HCN 1-0) which also shows contours of the column density for three levels: 8$\times10^{22}$, 2$\times10^{23}$, and 4$\times10^{23}$ cm$^{-2}$. For all of the maps, ratios are only computed where the column density is above a threshold of 7e22, and the molecular line is above a threshold of either 0.20 or 0.34 K \kms(see Section \ref{obs}).}
\label{ratio}
\end{figure}
\clearpage

\renewcommand\thefigure{2}
\begin{figure}
\includegraphics[scale=0.85]{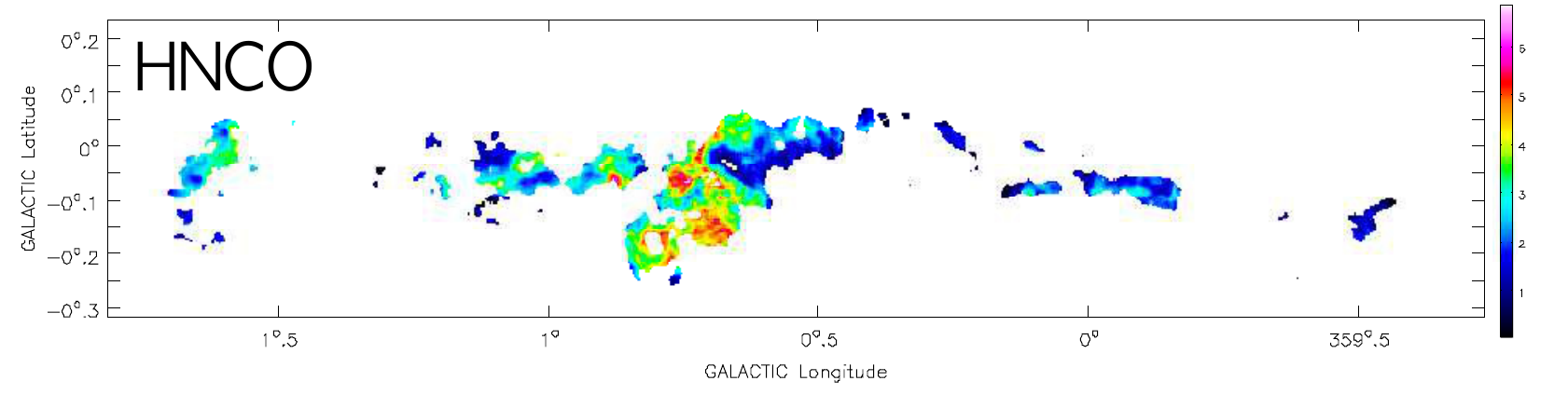}
\includegraphics[scale=0.85]{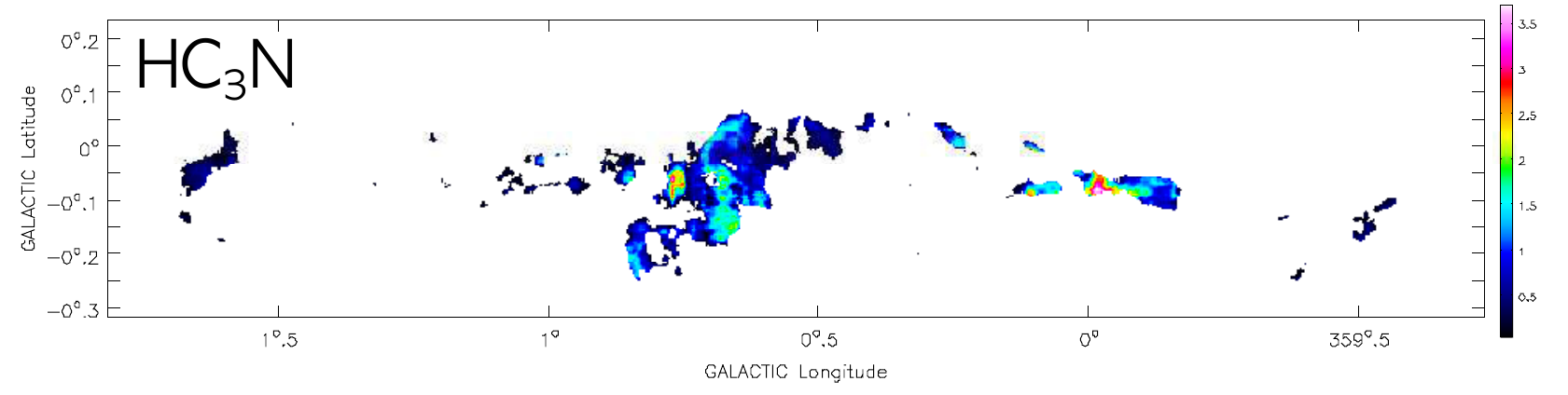}
\includegraphics[scale=0.85]{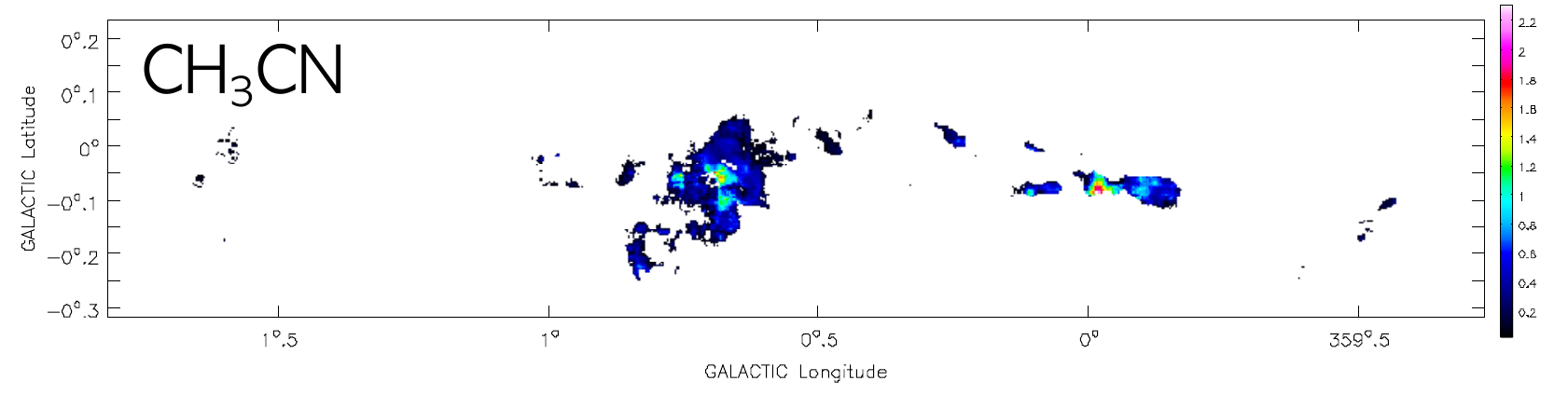}
\includegraphics[scale=0.85]{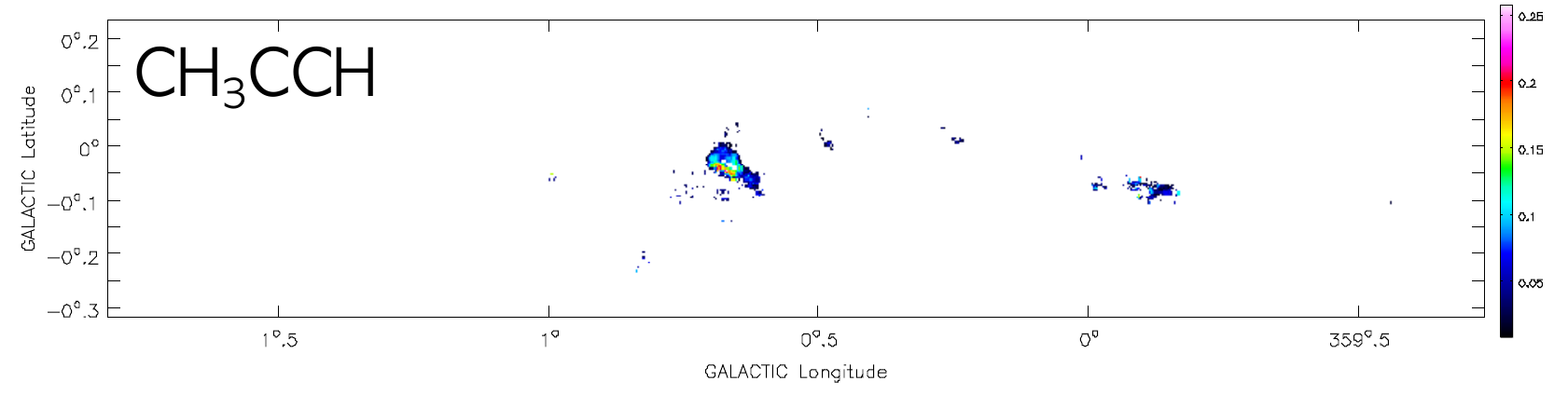}
\includegraphics[scale=0.85]{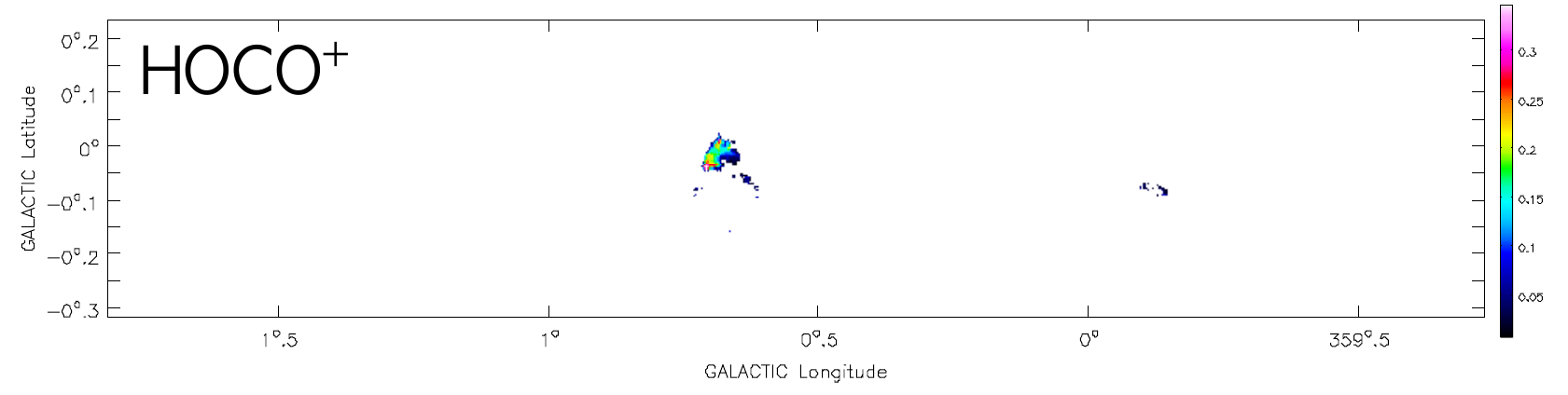}
\includegraphics[scale=0.85]{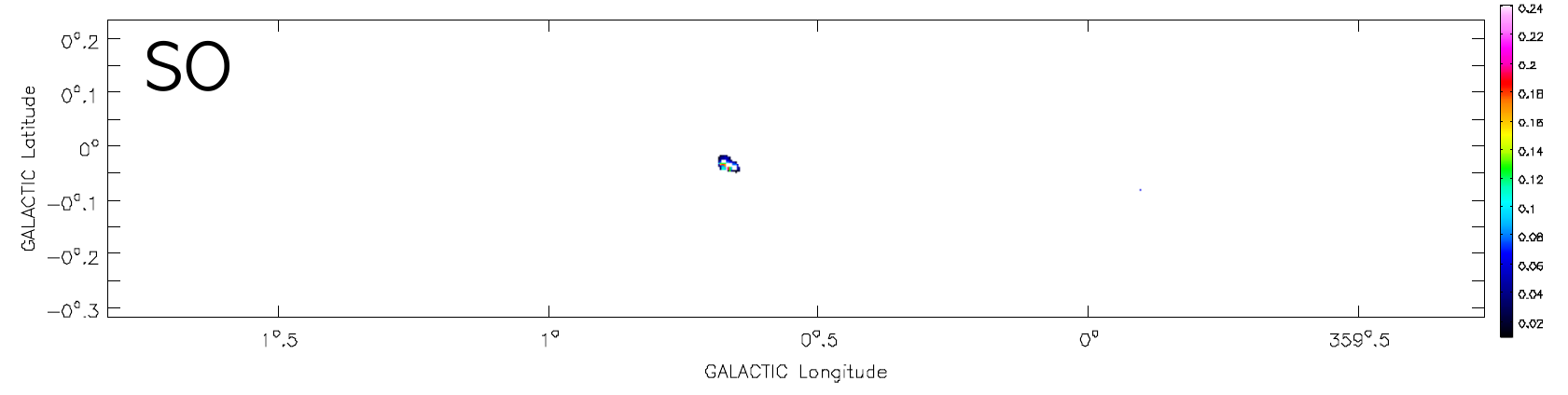}
\caption{Continued}
\end{figure}
\clearpage

\renewcommand\thefigure{2}
\begin{figure}
\includegraphics[scale=0.85]{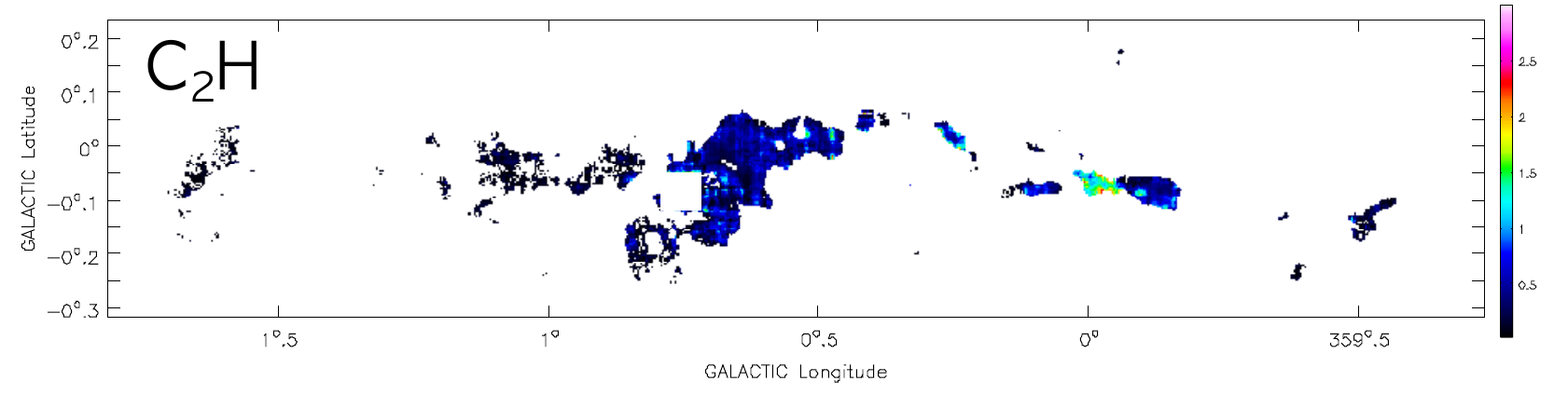}
\includegraphics[scale=0.85]{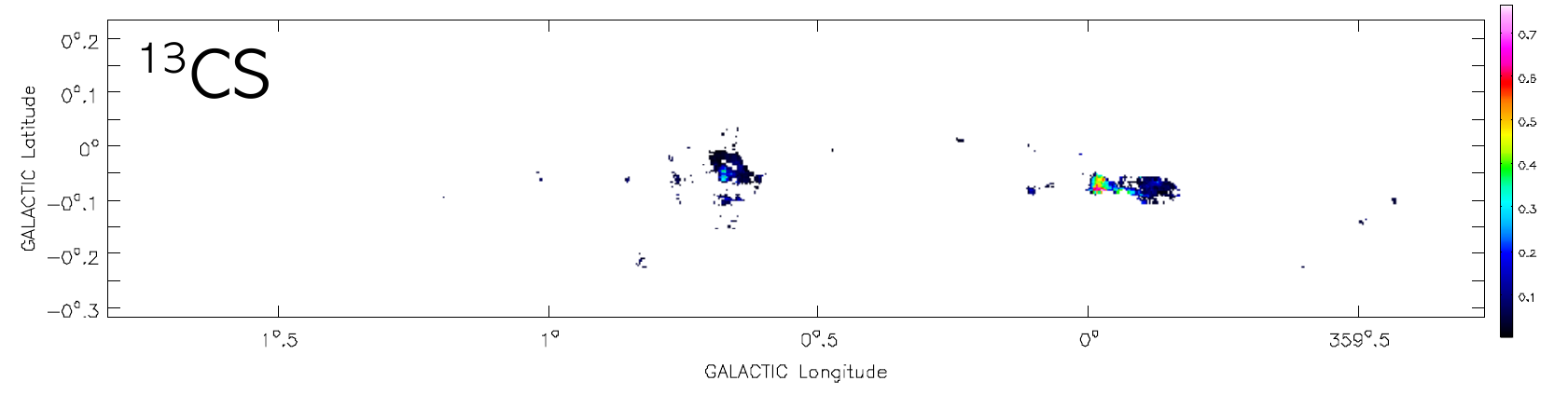}
\includegraphics[scale=0.85]{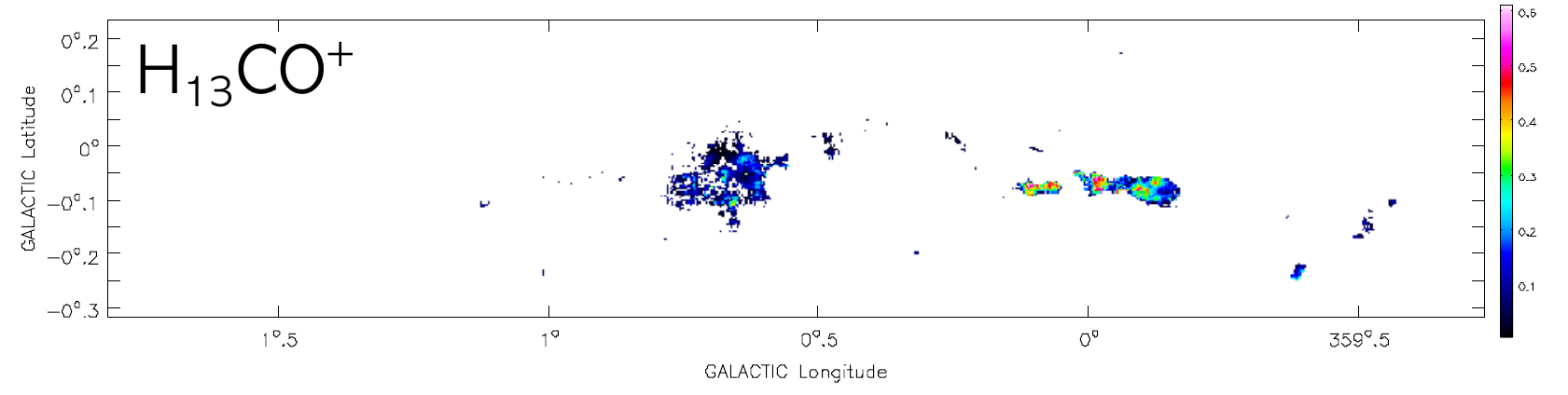}
\includegraphics[scale=0.85]{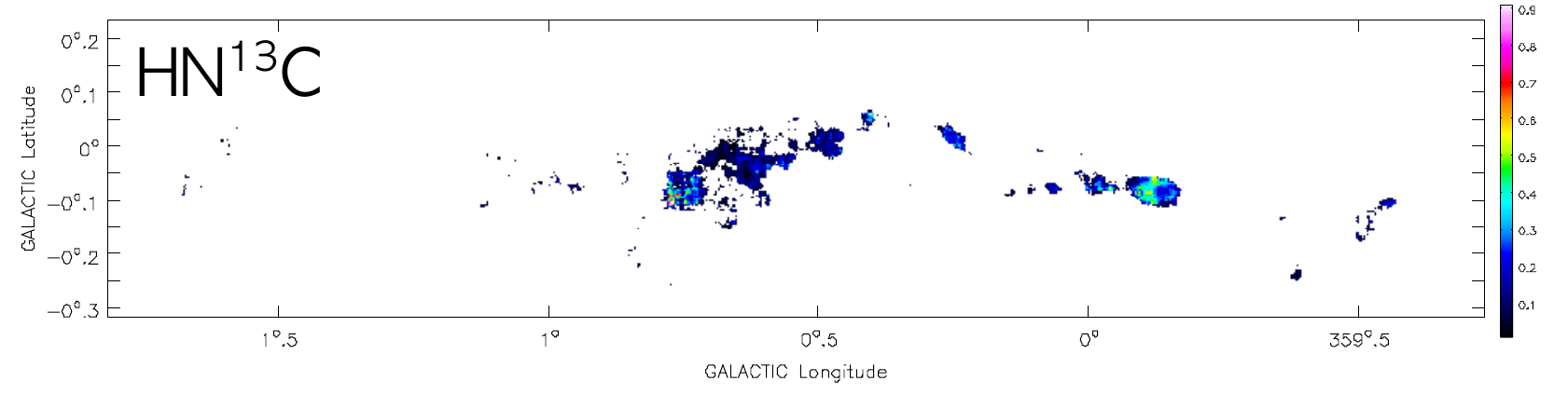}
\includegraphics[scale=0.85]{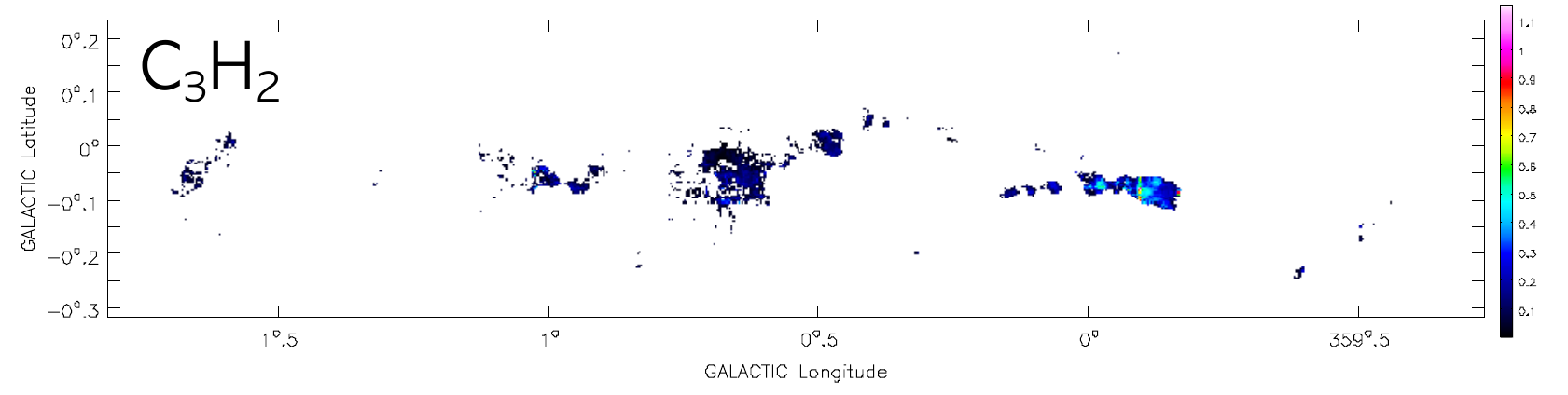}
\caption{Continued}
\end{figure}
\clearpage

\renewcommand\thefigure{3}
\begin{figure}
\hspace{-1.8cm}
\includegraphics[scale=0.6, angle=0]{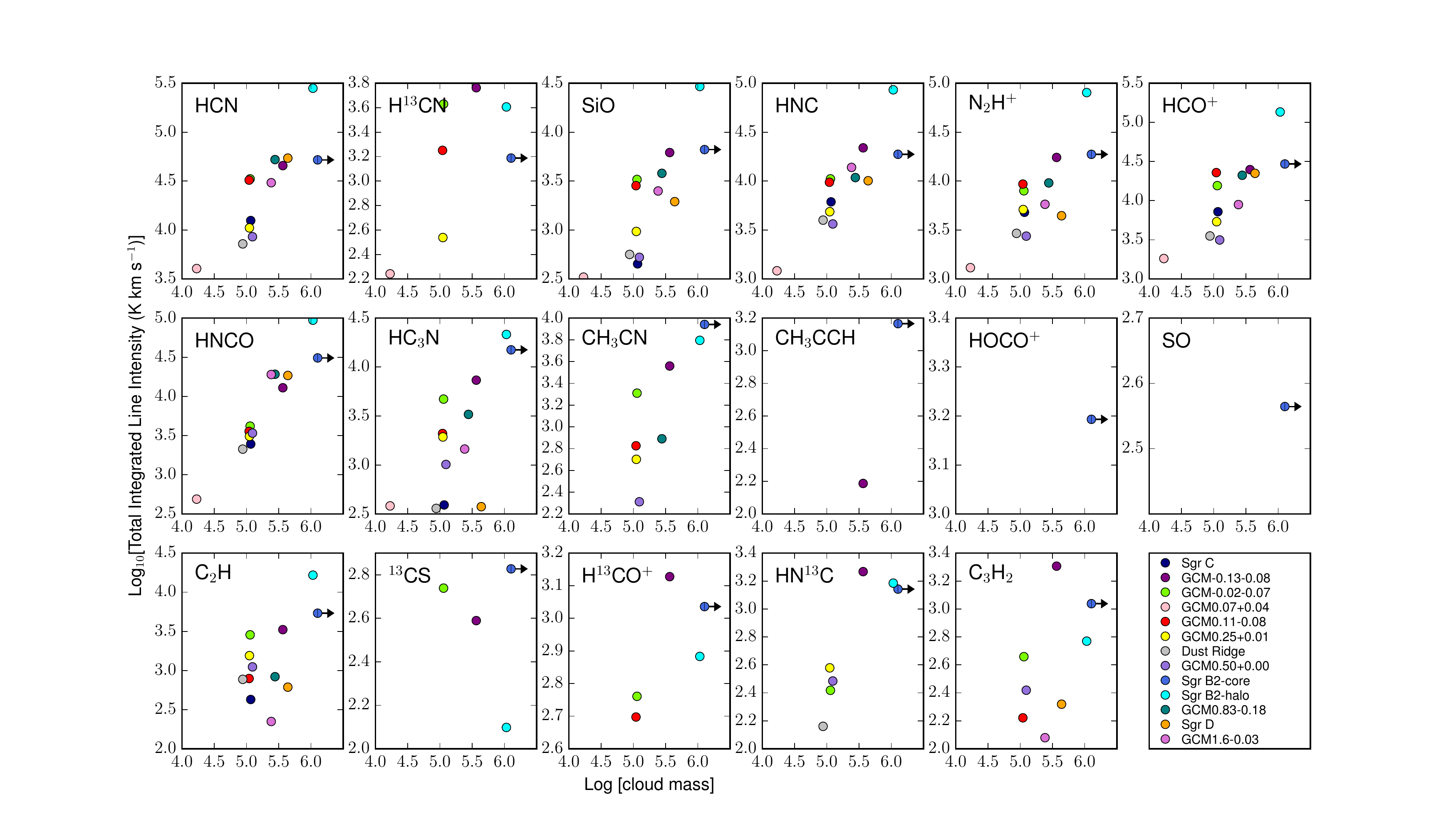}
\caption{Plots of the base-10 logarithm of the total intensity of HCN 1-0 and all other 3 mm lines from \cite{Jones12} compared to the base-10 logarithm of the total (dense) gas mass from the Herschel column density map, integrated over a sample of 13 Galactic center giant molecular clouds. The relevant uncertainties for this comparison are the errors in the relative calibrations across the molecular line and column density maps, and these are less than the size of the points plotted here. The mass of the Sgr B2 core is shown as a lower limit as the column density map of this source omits the brightest pixels at the peak of this cloud.}
\label{clouds}
\end{figure}
\clearpage

\renewcommand\thefigure{4}
\begin{figure}
\hspace{-1.8cm}
\includegraphics[scale=0.6, angle=0]{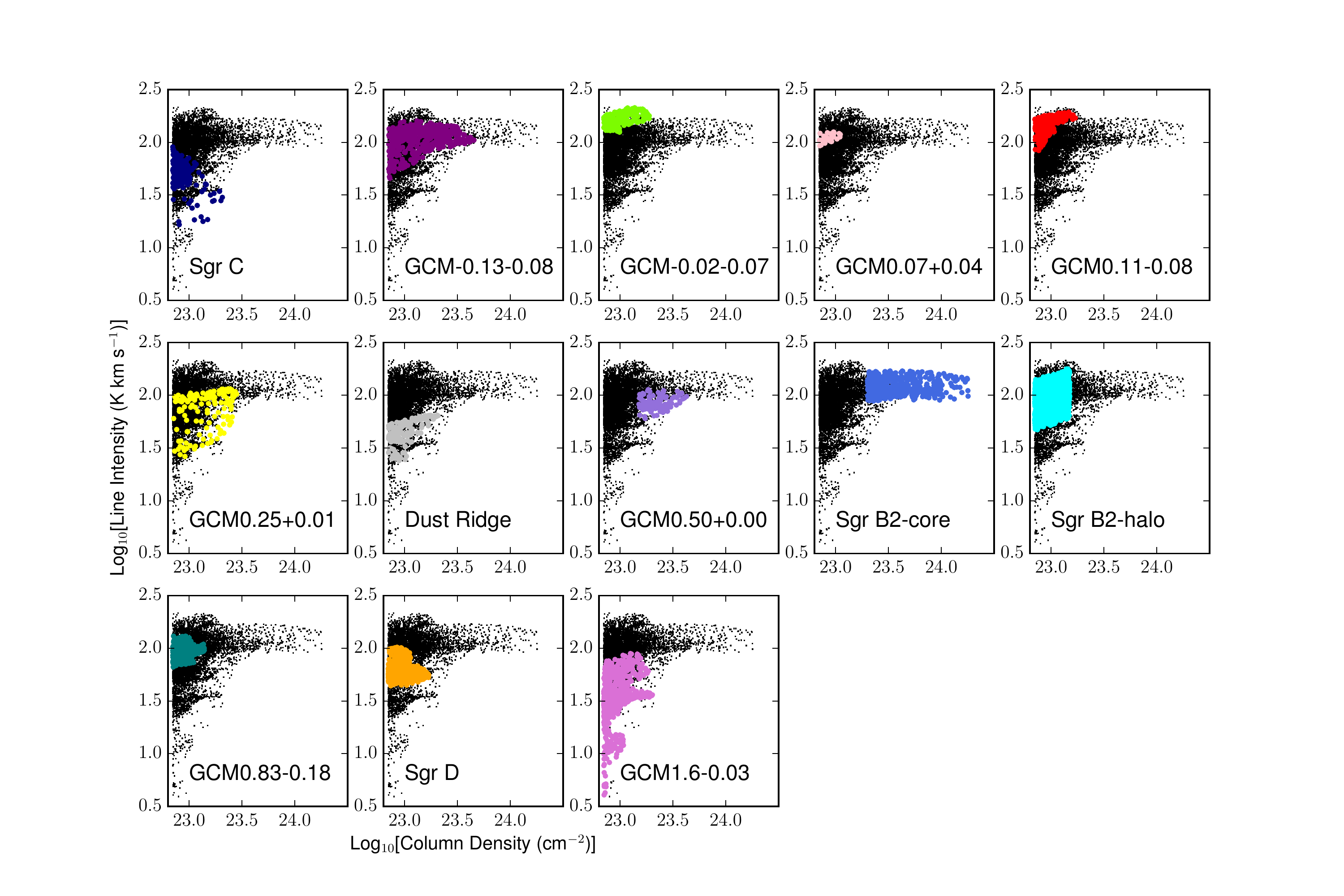}
\caption{Plots of the base-10 logarithm of the intensity of HCN 1-0 compared to the base-10 logarithm of the Herschel column density map on a pixel-by-pixel basis. Pixels corresponding to the 13 Galactic center giant molecular clouds analyzed in this paper are colorized according to the legend at the bottom right.}
\label{pixels}
\end{figure}
\clearpage

\renewcommand\thefigure{5}
\begin{figure}
\hspace{0.85cm}
\includegraphics[scale=0.181]{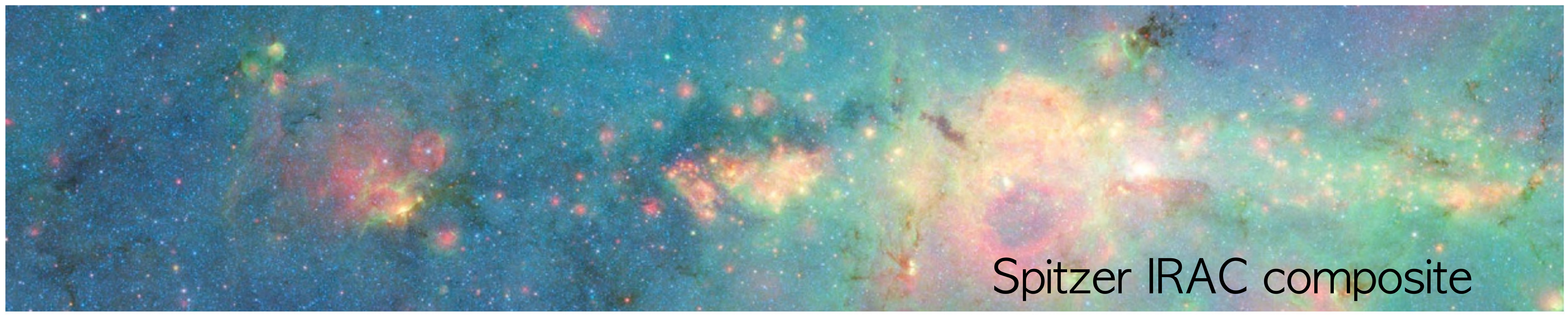}\\
\includegraphics[scale=0.85]{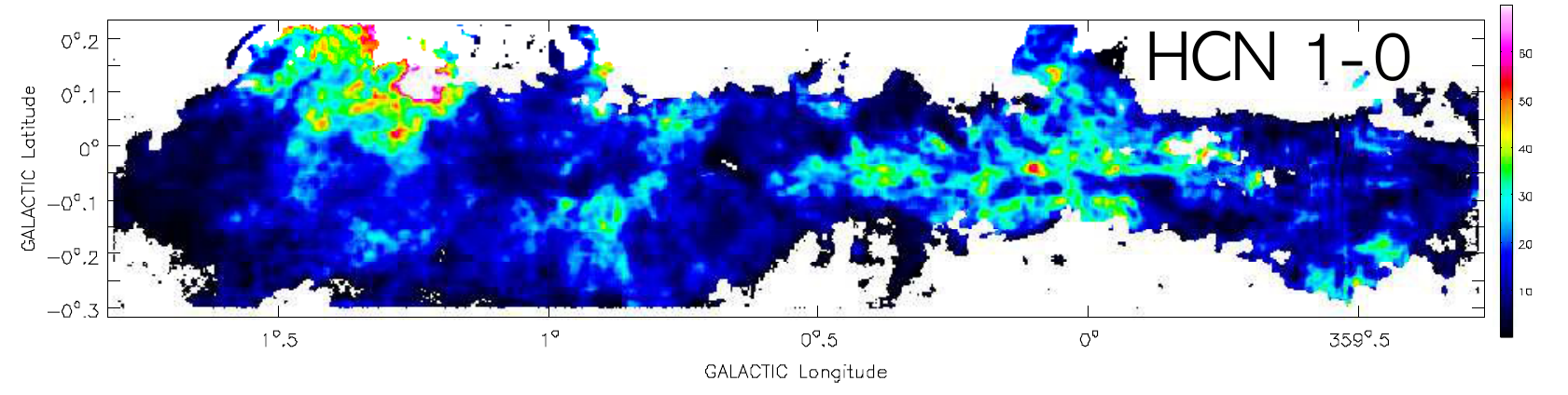}
\caption{{\bf Top:} A composite map of the infrared emission in the Galactic center from 3 to 8 microns, using Spitzer-IRAC. {\bf Bottom:} A map of the HCN 1-0 intensity to dense gas conversion factor. The units of the ratio shown are line intensity in K km/s over the column density in units of 10$^{22}$ cm$^{-2}$. The ratios is computed where the column density is above a threshold of 1$\times10^{22}$ cm$^{-2}$ and the HCN 1-0 intensity is above a threshold of 0.34 K \kms.}
\label{IR}
\end{figure}
\clearpage

\renewcommand\thefigure{6}
\begin{figure}
\hspace{0.85cm}
\includegraphics[scale=0.5]{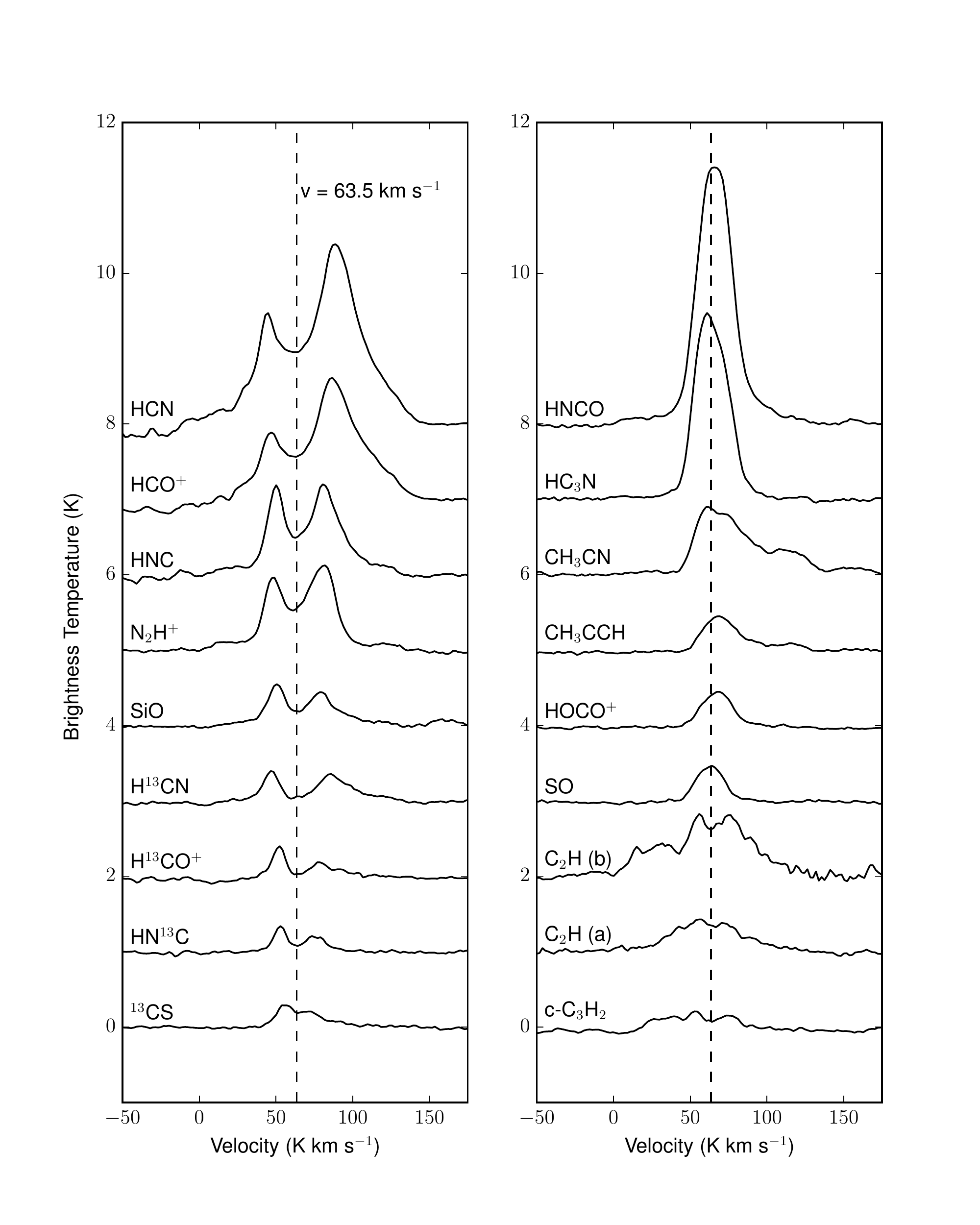}\\
\caption{Average spectra of emission toward the core of the Sgr B2 cloud. All species whose ground (1-0) transitions are observed exhibit a distinctive two-peak morphology, characteristic of self-absorption around a central velocity of $\sim$65 \kms. Note that the signature of self-absorption is seen even in the 1-0 transitions of less abundant $^{13}$C isotopologues. Self-absorption is also seen in the 2-1 transitions of \csiso\, SiO, and possibly \cthree, however the (2$_2$-1$_1$) transition of SO appears largely unaffected, as do transitions of more complex molecules with J$_{up}\ge$4.  }
\label{B2}
\end{figure}
\clearpage

\begin{table}[ht]
\caption{Linear fits to log (Intensity) vs log (Gas Mass) for Individual Clouds from a Single Species} 
\centering
\begin{tabular}{cccc}
\\[0.5ex]
\hline\hline
& & & \\
{\bf Species} & {\bf Slope} & {\bf r-value} & {\bf p-value} \\ [0.5ex]
\hline
HNCO & 1.18$\pm$0.11 & 0.95 &  $<$0.001 \\
HNC & 0.79$\pm$0.13 & 0.88 & $<$0.001 \\
HCN & 0.85$\pm$0.15 & 0.86 & $<$0.001 \\
HCO$^+$ & 0.85$\pm$0.17 & 0.84 & $<$0.001 \\
SiO & 0.93$\pm$0.20 & 0.82 & $<$0.001\\
N$_2$H$^+$ & 0.74$\pm$0.16 & 0.82 & $<$0.001 \\
\hline
HN$^{13}$C & 0.79$\pm$0.18 & 0.89 & 0.008 \\
CH$_3$CN & 1.08$\pm$0.29 & 0.84 &  0.01 \\
HC$_3$N & 0.86$\pm$0.27 & 0.69 & 0.008 \\
H$^{13}$CN & 0.58$\pm$0.30 & 0.66 & 0.1 \\
H$^{13}$CO$^+$ & 0.23$\pm$0.15 & 0.66 & 0.2 \\
C$_2$H & 0.77$\pm$0.33 & 0.59 &  0.04 \\
$^{13}$CS & -0.23$\pm$0.44 & -0.35 & 0.7 \\
C$_3$H$_2$ & 0.51$\pm$0.35 &  0.51 & 0.2 \\
\hline\hline
\end{tabular}
\label{stats}
\end{table}

\end{document}